\documentclass[12pt]{article}
\usepackage{amsmath,amssymb,amsfonts, amsbsy, epsfig, subfig}
\usepackage[usenames]{color}
\usepackage{rotating}

\newtheorem{theorem}{Theorem}
\newtheorem{corollary}{Corollary}
\newtheorem{proposition}{Proposition}
\newtheorem{lemma}{Lemma}
\newtheorem{example}{Example}
\newtheorem{remark}{Remark}
\newtheorem{definition}{Definition}
\newcommand{\beq}{\begin{equation}}
\newcommand{\eeq}{\end{equation}}
\newcommand{\beas}{\begin{eqnarray*}}
\newcommand{\eeas}{\end{eqnarray*}}
\newcommand{\bea}{\begin{eqnarray}}
\newcommand{\eea}{\end{eqnarray}}
\newcommand{\bei}{\begin{itemize}}
\newcommand{\eei}{\end{itemize}}
\newcommand{\ben}{\begin{enumerate}}
\newcommand{\een}{\end{enumerate}}
\newcommand{\bet}{\begin{theorem}}
\newcommand{\eet}{\end{theorem}}
\newcommand{\bel}{\begin{lemma}}
\newcommand{\eel}{\end{lemma}}
\newcommand{\bep}{\begin{proposition}}
\newcommand{\eep}{\end{proposition}}
\newcommand{\bed}{\begin{definition}}
\newcommand{\eed}{\end{definition}}
\newcommand{\bec}{\begin{corollary}}
\newcommand{\eec}{\end{corollary}}
\newcommand{\bex}{\begin{example}}
\newcommand{\eex}{\end{example}}
\newcommand{\qed}{\quad\hbox{\vrule width 4pt height 6pt depth 1.5pt}}
\newcommand{\RR}{I\!\! R}

\newcommand{\ep}{\epsilon}

\newcommand{\goto}{\rightarrow}
\newcommand{\hf}{{1 \over 2}}

\newcommand{\argmin}{\mathop{\rm arg\min}}

\newcommand{\sgn}{\mathop{\rm sgn}\nolimits}

\def\pr{\textsf{P}} 
\def\ep{\textsf{E}} 
\def\Cov{\textsf{Cov}} 
\def\Var{\textsf{Var}} 

\def\liminf{\mathop{\underline{\rm lim}}}

\begin{document}

\title{Adaptive Thresholding for Sparse Covariance Matrix Estimation}\author{Tony Cai$^{1}$ and Weidong Liu$^{1, 2}$}
\date{}

\maketitle

\footnotetext[1]{ Department of Statistics, The Wharton School, University of
  Pennsylvania,  Philadelphia, PA  \newline  \indent \ \
19104, tcai@wharton.upenn.edu. The research was supported in part
by NSF FRG  Grant DMS-0854973.}
\footnotetext[2]{Department of Mathematics and Institute of Natural Sciences, Shanghai Jiao Tong University,  China.}
 \vspace{-7mm}

\begin{abstract}
In this paper we consider  estimation of sparse covariance matrices and
propose a thresholding procedure which is adaptive to the variability
of individual entries. The estimators are fully data driven and
enjoy excellent performance both theoretically and
numerically. It is shown that the estimators adaptively achieve the
optimal rate of convergence over a large class of sparse covariance
matrices under the spectral norm. In
contrast, the commonly used universal thresholding estimators are shown to be
sub-optimal over the same parameter spaces. Support recovery is also discussed.
The adaptive thresholding estimators are easy to implement.
Numerical performance of the estimators is studied using both simulated and real data. Simulation results show that the adaptive thresholding estimators uniformly outperform the universal thresholding estimators. The method is also illustrated in an analysis on a dataset from
a small round blue-cell tumors microarray experiment. A supplement to this paper which contains additional technical proofs is available online.
\end{abstract}

\noindent{\bf Keywords:\/} Adaptive thresholding, Frobenius norm,
optimal rate of convergence, sparse covariance matrix, spectral norm,
support recovery, universal thresholding.

\newpage

\section{Introduction}
\label{intro.sec}

Let $\textbf{X}=(X_{1},\dotsc, X_{p})$ be a $p$-variate random
vector with covariance matrix $ \boldsymbol{\Sigma}_{0}$. Given an independent and
identically distributed (i.i.d.) random sample
$\{\textbf{X}_{1},\dotsc,\textbf{X}_{n}\}$ from the distribution of
$\textbf{X}$,  we wish to estimate the covariance matrix $\boldsymbol{\Sigma}_{0}$
under the spectral norm.
This covariance matrix estimation problem is of fundamental importance
in multivariate analysis with a wide range of applications.
The high dimensional setting, where the dimension $p$ can be much larger
than the sample size $n$, is of particular current interest. In such a setting,
conventional methods and results based on fixed $p$ and large $n$ are
no longer applicable and new methods and theories are thus needed.
In particular, the sample covariance matrix
\beq
\label{sample.covariance}
  \boldsymbol{\Sigma}_{n}=(\hat \sigma_{ij})_{p\times p}:=\frac{1}{n-1}\sum_{k=1}^{n}(\textbf{X}_{k}-\bar{\textbf{X}})(\textbf{X}_{k}-\bar{\textbf{X}})^{T},
\eeq
where $\bar{\textbf{X}}=n^{-1}\sum_{k=1}^{n}\textbf{X}_{k}$, performs poorly
in this setting and structural assumptions are required in order to estimate the covariance
matrix consistently.

In this paper we focus on estimating sparse
covariance matrices. This problem has been considered in the literature.
El Karoui (2008) and Bickel and Levina (2008) proposed thresholding
of the sample covariance matrix $ \boldsymbol{\Sigma}_n$ and obtained rates of
convergence for the thresholding estimators. Rothman, Levina and Zhu
(2009) considered thresholding of the sample covariance matrix with
more general thresholding functions. Cai and Zhou (2009 and 2010)
established the minimax rates of convergence under the matrix
$\ell_{1}$ norm and the spectral norm. Wang and Zou (2010)
considered estimation of volatility matrices based on high-frequency
financial data.

A common feature of the thresholding methods for sparse covariance
matrix estimation proposed in the literature so far is that they
all belong to the class of ``universal thresholding rules''. That
is, a single threshold level is used to threshold all the entries of
the sample covariance matrix. Universal thresholding rules were
originally introduced by Donoho and Johnstone (1994 and 1998) for
estimating sparse normal mean vectors in the context of wavelet
function estimation. See also Antoniadis and Fan (2001). An important feature of problems considered
there is that noise is homoscedastic. In such a setting, universal
thresholding has demonstrated considerable success in nonparametric
function estimation in terms of asymptotic optimality and
computational simplicity.

In contrast to the standard homoscedastic nonparametric regression
problems, sparse covariance matrix estimation is intrinsically a
heteroscedastic problem in the sense that the entries
of the sample covariance matrix could have a wide range of
variability. Although some universal thresholding rules have been
shown to enjoy desirable asymptotic properties,  this is
mainly due to the fact that the parameter space
considered in the literature is relatively  restrictive which
forces the covariance matrix estimation problem to be an
essentially homoscedastic problem.

To illustrate the point, it is helpful to consider an idealized model where one observes
\beq
\label{mean.model}
y_i = \mu_i + \gamma_i z_i, \quad z_i\stackrel{iid}{\sim} N(0, 1)\quad 1\le i \le p
\eeq
and wishes to estimate the mean vector $\mu$ which is assumed to be sparse. If  the noise levels $\gamma_i$ are bounded, say by $B$, then the
universal thresholding rule $\hat \mu_i = y_i I(|y_i| \ge B
\sqrt{2\log p})$ performs well asymptotically over a standard $\ell_q$ ball $\Theta_q(s_0)$ defined by
\beq
\Theta_q(s_0) = \{ \boldsymbol{\mu}\in \RR^p: \sum_{j=1}^{p}|\mu_{j}|^{q}\leq s_{0}\}.
\eeq
In particular, $\Theta_0(s_0)$ is a set of sparse vectors with at most $s_{0}$ nonzero elements.
Here the assumption that
$\gamma_i$ are bounded by $B$ is crucial. The universal thresholding
rule simply treats the heteroscedastic problem (\ref{mean.model}) as a
homoscedastic one with all noise levels $\gamma_i = B$. It is
intuitively clear that this method does not perform well when the range
of $\gamma_i$ is large and it fails completely without the uniform
boundedness assumption on the $\gamma_i$'s.

For sparse covariance matrix estimation, the following uniformity
class of sparse matrices was considered in Bickel and Levina (2008)
and Rothman, Levina and Zhu (2009):
\begin{eqnarray*}
\mathcal{U}_{q}:=\mathcal{U}_{q}(s_{0}(p))=\Big{\{} \boldsymbol{\Sigma}:  \boldsymbol{\Sigma}\succ 0, ~
\max_{i}\sigma_{ii}\leq K, ~ \max_{i}\sum_{j=1}^{p}|\sigma_{ij}|^{q}\leq s_{0}(p)\Big{\}}
\end{eqnarray*}
for some $0\leq q<1$, where  $ \boldsymbol{\Sigma}\succ 0$ means that $ \boldsymbol{\Sigma}$ is positive
definite.  Here each column of a covariance matrix in
$\mathcal{U}_{q}(s_{0}(p))$ is assumed to be in the $\ell_q$ ball
$\Theta_q(s_0(p))$. Define
\beq
\label{thetaij}
\theta_{ij} :=\Var((X_{i}-\mu_{i})(X_{j}-\mu_{j})),
\eeq
where $\mu_{i}=\ep X_{i}$. It is easy to see that in the Gaussian case,
$\sigma_{ii}\sigma_{jj}\le \theta_{ij}\le 2\sigma_{ii}\sigma_{jj}$.
The condition $\max_{i}\sigma_{ii}\le K$ for all $i$ ensures the variances of the entries of the sample covariance matrix to be uniformly bounded.
Bickel and Levina (2008) proposed a universal thresholding estimator
$ \hat{\boldsymbol{\Sigma}}_{u}=(\hat{\sigma}^{u}_{ij})$, where
\begin{eqnarray}\label{a1}
\hat{\sigma}^{u}_{ij}=\hat{\sigma}_{ij}I\{\hat{\sigma}_{ij}\geq
\lambda_{n}\},
\end{eqnarray}
and showed that with a proper choice of the threshold $\lambda_n$ the
estimator $\hat{\boldsymbol{\Sigma}}_{u}$ achieves a desirable rate of
convergence under the spectral norm.
Rothman, Levina and Zhu (2009) considered a class of universal
thresholding rules with more general thresholding functions than hard thresholding.
Similar to the idealized model (\ref{mean.model}) discussed earlier, here a key assumption is that the variances $\sigma_{ii}$ are uniformly bounded by $K$ which is crucial to make the universal thresholding rules well behaved. A universal thresholding rule in this case essentially treats the problem as if all $\sigma_{ii} =K$ when selects the threshold $\lambda$.

For heteroscedastic problems such as sparse covariance matrix
estimation, it is arguable more desirable to use thresholds that capture the
variability of individual variables instead of using a universal upper
bound. This is particularly true when the variances vary over a wide range or no obvious upper bound on the variances
is known. A more natural and effective approach is to use
thresholding rules with entry-dependent thresholds which automatically
adapt to the variability of the individual entries of the sample covariance
matrix. The main goal of the present paper is to develop such an adaptive
thresholding estimator and study its properties.

In this paper we introduce an adaptive thresholding estimator
$\hat{\boldsymbol{\Sigma}}^{\star}=(\hat{\sigma}^{\star}_{ij})_{p\times p}$ with
\beq
\label{estimator}
\hat{\sigma}^{\star}_{ij}=s_{\lambda_{ij}}(\hat{\sigma}_{ij}),
\eeq
where $s_{\lambda}(z)$ is a general thresholding function similar to
those used in Rothman, Levina and Zhu (2009) and will be specified
later. The individual thresholds $\lambda_{ij}$ are fully
data-driven and adapt to the variability of individual entries of
the sample covariance matrix $\boldsymbol{\Sigma}_n$. It is shown that the
adaptive thresholding estimator $\hat{\boldsymbol{\Sigma}}^{\star}$ enjoys
excellent properties both asymptotically and numerically. In
particular, we consider the performance of the estimator
$\hat{\boldsymbol{\Sigma}}^{\star}$ over a large class of sparse covariance
matrices defined by
\beq
\label{cov.space}
\mathcal{U}^{\star}_{q}:=\mathcal{U}^{\star}_{q}(s_{0}(p))=\Big{\{}\boldsymbol{\Sigma}: \boldsymbol{\Sigma}\succ 0, ~
\max_{i}\sum_{j=1}^{p}(\sigma_{ii}\sigma_{jj})^{(1-q)/2}|\sigma_{ij}|^{q}\leq
s_{0}(p)\Big{\}}
\eeq
for $0\leq q<1$. In comparison to $\mathcal{U}_{q}(s_{0}(p))$, the columns of a covariance matrix in
$\mathcal{U}^{\star}_{q}$ are required to be in a weighted $\ell_q$
ball, instead of a standard $\ell_q$ ball, with the weight determined by the
variance of the entries of the sample covariance matrix.
A particular feature of $\mathcal{U}^{\star}_{q}$ is that it no longer
requires the variances $\sigma_{ii}$ to be uniformly bounded and allows
$\max_{i}\sigma_{ii}\rightarrow\infty$.
Note that $\mathcal{U}_{q}(s_{0}(p))\subseteq
\mathcal{U}^{\star}_{q}(K^{1-q}s_{0}(p))$, so the parameter space
$\mathcal{U}^{\star}_{q}$ contains the uniformity class
$\mathcal{U}_{q}$ as a subset. The parameter space $\mathcal{U}^{\star}_{q}$ can also be viewed as a weighted $\ell_q$ ball of correlation coefficients. See Section \ref{rate.sec} for more discussions.

It will be shown in Section \ref{theory.sec} that  $\hat{\boldsymbol{\Sigma}}^{\star}$ achieves the optimal rate of convergence
\[
s_0(p) \left({\log p\over n}\right)^{(1-q)/2}
\]
over the parameter space $\mathcal{U}^{\star}_{q}(s_{0}(p))$.
In comparison, it is also shown that the best universal thresholding
estimator can only attain the rate
$
s_0^{2-q}(p) \left({\log p\over n}\right)^{(1-q)/2}
$
over $\mathcal{U}^{\star}_{q}(s_{0}(p))$, which is clearly sub-optimal  when $s_{0}(p)\rightarrow\infty$
since $q<1$.

The choice of the regularization parameters is important in any
regularized estimation problem. The thresholds $\lambda_{ij}$ used in
(\ref{estimator}) are based on
an estimator of the variance of the entries
$\hat \sigma_{ij}$ of the sample covariance matrix. More specifically,
$\lambda_{ij}$  are of the form
\beq
\label{lambdaij}
\lambda_{ij} = \delta \sqrt{\hat \theta_{ij}\log p \over n}
\eeq
where $\hat \theta_{ij}$ are estimates of $\theta_{ij}$ defined in
(\ref{thetaij}) and $\delta$ is a tuning parameter. The value of
$\delta$ can be taken as
fixed at $\delta=2$, or it can be empirically chosen through cross
validation.  Theoretical properties of
the resulting covariance matrix estimators using both methods are
investigated. It is shown that the estimators attain the optimal rate
of convergence under the spectral norm in both cases.
In addition,  support recovery of a sparse covariance matrix is also
considered.

The adaptive thresholding estimators are easy to implement.
Numerical performance of the estimators is investigated using both
simulated and real data.  Simulation results show that the
adaptive thresholding estimators perform favorably in comparison to
existing methods. In particular, they uniformly outperform the
universal thresholding estimators in the simulation studies.
The procedure is also applied to analyze a dataset from a small round
blue-cell tumors  microarray experiment (Khan et al., 2001).

The paper is organized as follows. Section
\ref{method.sec} introduces the adaptive thresholding
procedure for sparse covariance matrix estimation. Asymptotic
properties are studied in Section \ref{theory.sec}. It is shown that
the adaptive thresholding estimator is rate-optimal over
$\mathcal{U}^{\star}_{q}$, while the best universal thresholding
estimator is proved to be suboptimal. Section \ref{delta.sec} discusses
data-driven selection of the thresholds using cross validation and
establish asymptotic optimality of the resulting estimator.
Numerical performance of the adaptive thresholding estimators is
investigated by simulations and by an
application to a dataset from a small round blue-cell tumors  microarray
experiment in Section \ref{numerical.sec}.
Section \ref{discussion.sec} discusses methods based on the sample correlation matrix. The proofs
are given in Section \ref{proof.sec}.

\section{Adaptive thresholding for sparse covariance matrix}
\label{method.sec}

In this section we introduce the adaptive thresholding method for
estimating sparse covariance matrices.
To motivate our estimator, consider again
the sparse normal mean estimation problem (\ref{mean.model}). If the
noise levels $\gamma_i$'s are known or can be well estimated, a
good estimator of the mean vector  is the hard thresholding
estimator $\hat{\mu}_{i}=y_{i}I\{|y_{i}|\geq \gamma_{i}\sqrt{2\log
p}\}$ or some generalized thresholding estimator with the same thresholds
$\gamma_{i}\sqrt{2\log p}$.

Similarly, for sparse covariance matrix estimation, a more effective
thresholding rule than universal thresholding is the one which adapts
to the variability of the
individual entries of the sample covariance matrix. Define $\theta_{ij}$
as in (\ref{thetaij}).
Then, roughly speaking, estimation of a sparse covariance matrix is
similar to the mean vector estimation problem based on the
observations \beq \label{mean.model2} \frac{1}{n}\sum_{k=1}^{n}
(X_{ki}-\mu_{i})(X_{kj}-\mu_{j})=\sigma_{ij}+\sqrt{\theta_{ij}\over
n} z_{ij}, \quad 1\le i, \; j \le p \eeq with $z_{ij}$ being
asymptotically standard normal. This analogy provides a good
motivation for our adaptive thresholding procedure. If the
$\theta_{ij}$ were known, a natural thresholding estimator would be
$(\hat{\sigma}^{o}_{ij})_{p\times p}$ with
\beq
\label{est0}
\hat{\sigma}^{o}_{ij}=s_{\lambda^{o}_{ij}}(\hat{\sigma}_{ij}) \quad
\mbox{with~}\lambda^{o}_{ij}=2 \sqrt{\theta_{ij}\log p\over n},
\eeq
where $s_{\lambda}(z)$ is a  thresholding function.  Comparing to
the universal thresholding rule in Bickel and Levina (2008), the
variance factors $\theta_{ij}$ in the thresholds make the
thresholding rule entry-dependent and leads to a more flexible
estimator.  In practice, $\theta_{ij}$ are typically unknown, but
can be well estimated. We propose the following estimator of
$\theta_{ij}$:
\begin{eqnarray*}
\hat{\theta}_{ij}=\frac{1}{n}\sum_{k=1}^{n}\Big{[}
(X_{ki}-\bar{X}^{i})(X_{kj}-\bar{X}^{j})-\hat{\sigma}_{ij}\Big{]}^{2},\quad
\bar{X}^{i}=n^{-1}\sum_{k=1}^{n}X_{ki}.
\end{eqnarray*}
This leads to our adaptive thresholding estimator of the covariance matrix $\boldsymbol{\Sigma}_0$,
\begin{eqnarray}\label{adap}
\hat{\boldsymbol{\Sigma}}^{\star}(\delta)=(\hat{\sigma}^{\star}_{ij})_{p\times
p}\quad \mbox{with}\quad
\hat{\sigma}^{\star}_{ij}=s_{\lambda_{ij}}(\hat{\sigma}_{ij}),
\end{eqnarray}
where
\begin{eqnarray}\label{c1}
\lambda_{ij}:=\lambda_{ij}(\delta)=\delta\sqrt{\frac{\hat{\theta}_{ij}\log
 p}{n}}.
 \end{eqnarray}
Here $\delta > 0$ is a regularization parameter. It can be
fixed at $\delta = 2$ or can be chosen through cross validation. Good choices of $\delta$ will not affect the rate of convergence, but will affect the numerical performance of the resulting estimators. Selection of $\delta$ is thus of practical importance and we will discuss it further later.

The analogy between the sparse covariance estimation problem and the
idealized mean estimation problem (\ref{mean.model2}) gives a good
motivation for the adaptive thresholding estimator defined in
(\ref{adap}) and (\ref{c1}), but of course the matrix estimation
problem is not exactly equivalent to the mean estimation problem
(\ref{mean.model2}) as noise is not exactly normal or iid and the loss
is the spectral norm, not a vector norm or the Frobenius norm. We shall make  our technical analysis precise in Sections \ref{theory.sec} and \ref{proof.sec}.

In the present paper, we consider simultaneously a class of
thresholding functions $s_{\lambda}(z)$ that satisfy the following conditions:

\bei
\item[(i).] $|s_{\lambda}(z)|\leq c|y|$ for all $z,y$ satisfy $|z-y|\leq \lambda$ and some $c>0$;
\item[(ii).] $s_{\lambda}(z)=0$ for $|z|\leq \lambda$;
\item[(iii).] $|s_{\lambda}(z)-z|\leq \lambda$, for all $z\in\RR$.
\eei
These three conditions are satisfied, for example, by the soft
thresholding rule $s_{\lambda}(z)=\sgn(z)(z-\lambda)_{+}$ and the
adaptive lasso rule $s_{\lambda}(z)=z(1-|\lambda/z|^{\eta})_+$ with
$\eta\geq 1$, as called in Rothman, Levina and Zhu (2009). We
shall present a unified analysis of the adaptive thresholding
estimators with the thresholding function $s_{\lambda}(z)$ satisfying
the above three conditions. It should be noted that Condition (i)
excludes the hard thresholding rule.  However, all of the theoretical
results in this paper hold for the hard thresholding estimator under
similar conditions. Here Condition (i) is in place only to make the
technical analysis in Section \ref{proof.sec} work in a unified way for the class of
thresholding rules. The results for the hard
thresholding rule require slightly different proofs.

Rothman, Levina and Zhu (2009) proposed generalized universal
thresholding estimators
\begin{eqnarray*}
\hat{\boldsymbol{\Sigma}}_{g}=(\hat{\sigma}^{g}_{ij})_{p\times p},
\mbox{~~where~~}
\hat{\sigma}^{g}_{ij}=\bar{s}_{\lambda_{n}}(\hat{\sigma}_{ij})
\end{eqnarray*}
 and
$\bar{s}_{\lambda}(z)$ satisfies (ii), (iii) and $|\bar{s}_{\lambda}(z)|\leq |z|$, which is slightly weaker than (i).
Similar general universal thresholding rules were introduced and studied by Antoniadis and Fan (2001)  in the context of wavelet function estimation.
We should note that the generalized universal thresholding
estimators $\hat{\boldsymbol{\Sigma}}_{g}$ suffer the same shortcomings as those
of $\hat{\boldsymbol{\Sigma}}_{u}$, and like $\hat{\boldsymbol{\Sigma}}_{u}$ they are
sub-optimal over the class $\mathcal{U}^{\star}_{q}$.


\section{Theoretical properties of adaptive thresholding}
\label{theory.sec}

We now consider the asymptotic properties of the adaptive thresholding
estimator $\hat{\boldsymbol{\Sigma}}^{\star}(\delta)$ defined in (\ref{adap}) and
(\ref{c1}). It is shown that the estimator $\hat{\boldsymbol{\Sigma}}^{\star}(\delta)$ adaptively attains the optimal rate of convergence over the collection of parameter spaces $\mathcal{U}^{\star}_{q}(s_0(p))$.

We begin with some notation. Define the standardized variables
\[
Y_{i}=(X_{i}-\mu_{i})/(\Var(X_{i}))^{1/2},
\]
where $\mu_{i}=\ep X_{i}$, and let $\textbf{Y}=(Y_{1},\ldots,Y_{p}).$ Throughout the paper, denote
$|\textbf{a}|_{2}=\sqrt{\sum_{j=1}^{p}a^{2}_{j}}$ for the usual
Euclidean norm of a vector
$\textbf{a}=(a_{1},\dotsc,a_{p})^{T}\in \RR^{p}$.  For a matrix
$\boldsymbol{A}=(a_{ij})\in\RR^{p\times q}$, define  the spectral norm
$\|\boldsymbol{A}\|_{2}=\sup_{|\textbf{x}|_{2}\leq
  1}|\boldsymbol{A}\textbf{x}|_{2}$, the matrix $\ell_1$ norm $\|\boldsymbol{A}\|_{L_{1}}=\max_{1\leq
  j\leq q}\sum_{i=1}^{p}|a_{i,j}|$, and the Frobenius norm
$\|\boldsymbol{A}\|_{F}=\sqrt{\sum_{i,j}a^{2}_{ij}}$.  For two  sequences of real numbers
$\{a_{n}\}$ and $\{b_{n}\}$, write $a_{n} = O(b_{n})$ if there
exists a constant $C$ such that $|a_{n}| \leq C|b_{n}|$ holds for all
sufficiently large $n$, and write $a_{n} = o(b_{n})$ if
$\lim_{n\rightarrow\infty}a_{n}/b_{n} = 0$.

\subsection{Rate of convergence}
\label{rate.sec}

It is conventional in the covariance matrix estimation literature to
divide the technical analysis into
two cases according the the moment conditions on $\textbf{X}$.

 {\bf (C1). (Exponential-type tails)}  Suppose that $\log p= o(n^{1/3})$ and there exists some
 $\eta>0$ such that
\begin{eqnarray}\label{c3-4}
\ep \exp\Big{(}tY^{2}_{i}\Big{)}\leq K_{1}<\infty~~~\mbox{for all
$|t|\leq \eta$ and $i$.}
\end{eqnarray}
Furthermore, we assume that for some $\tau_{0}>0$,
\begin{eqnarray}\label{c2}
\min_{ij}\Var(Y_{i}Y_{j})\geq \tau_{0}.
\end{eqnarray}

{\bf (C2). (Polynomial-type tails)} Suppose that for some $\gamma,
c_{1}>0$,  $ p\leq c_{1}n^{\gamma}$, and for some $\epsilon>0$
\begin{eqnarray}\label{c3-3}
\ep|Y_{i}|^{4\gamma+4+\epsilon}\leq K_{1}~~~\mbox{for all $i$.}
\end{eqnarray}
Furthermore, we assume that (\ref{c2}) holds.\\\vspace{-3mm}

\begin{remark}
{\rm  Note that (C1) holds with $\tau_0 =1$ in the Gaussian case where $\textbf{X}\sim
N(\boldsymbol{\mu},\boldsymbol{\Sigma}_{0})$. To this end, let $\rho_{ij}$ be the correlation
coefficient of $Y_{i}$ and $Y_{j}$. We can then write
$Y_{i}=\rho_{ij}Y_{j} + \sqrt{1-\rho_{ij}^{2}}Y$, where $Y\sim N(0,1)$
is independent of $Y_{j}$. So we have
$\Var(Y_{i}Y_{j})=1+\rho^{2}_{ij}\geq 1$. Hence (\ref{c2}) holds
with $\tau_{0}=1$.
}
\end{remark}


The follow theorem gives the rate of convergence over the parameter space $\mathcal{U}^{\star}_{q}$ under the spectral norm for the thresholding estimator $\hat{\boldsymbol{\Sigma}}^{\star}(\delta)$.

\begin{theorem}
\label{bpsth4} Let $\delta\geq 2$ and $0\leq q<1$.
\bei
\item[{\rm (i).}] Under (C1),  we have, for some constant $C_{K_{1},\delta,c,q}$ depending only on
 $\delta$, $c$, $q$ and $K_{1}$,
\beq
\label{thre-1}
\inf_{\boldsymbol{\Sigma}_{0}\in\mathcal{U}^{\star}_{q}}\pr\Big{(}\|\hat{\boldsymbol{\Sigma}}^{\star}(\delta)-\boldsymbol{\Sigma}_{0}\|_{2}\leq
C_{K_{1},\delta,c,q}s_{0}(p)\Big{(}\frac{\log
p}{n}\Big{)}^{1-q\over 2}\Big{)}\geq 1-O((\log p)^{-\hf}p^{-\delta+2}).
\eeq

\item[{\rm (ii).}] Under (C2), (\ref{thre-1}) holds with  probability
greater than $1-O((\log p)^{-1/2}p^{-\delta+2}+n^{-\epsilon/8})$.
\eei
\end{theorem}

Although $\mathcal{U}^{\star}_{q}$ is larger than the uniformity
class $\mathcal{U}_{q}$, the rates of convergence of
$\hat{\boldsymbol{\Sigma}}^{\star}(\delta)$ over the two classes are of the same
order $s_{0}(p)(\log p/n)^{(1-q)/2}$.

Theorem \ref{bpsth4} states the rate of convergence in terms of
probability. The same rate of convergence holds in expectation with
some additional mild assumptions. By (\ref{thre-1}) and some long
but elementary calculations (see also the proof of Lemma \ref{le4}),
we have the following result on the mean squared spectral norm.

 \bep
 Under (C1) and $p\geq n^{\xi}$ for some $\xi>0$, we have  for $\delta\geq 7+\xi^{-1}$, $0\leq
 q<1$, and some  constant $C>0$,
\beq
\label{rate.expectation}
\sup_{\boldsymbol{\Sigma}_{0}\in\mathcal{U}^{\star}_{q}}\ep\|\hat{\boldsymbol{\Sigma}}^{\star}(\delta)-\boldsymbol{\Sigma}_{0}\|^{2}_{2}\leq
Cs^{2}_{0}(p)\Big{(}\frac{\log p}{n}\Big{)}^{1-q}.
\eeq
\eep

\begin{remark}
{\rm   Cai and Zhou (2010) established the minimax rates of convergence
under the spectral norm for sparse covariance matrix estimation over
$\mathcal{U}_{q}$. It was shown that the optimal rate over
$\mathcal{U}_{q}$ is $s_{0}(p)(\log p/n)^{(1-q)/2}$.  Since
$\mathcal{U}_{q}(s_{0}(p))\subseteq
\mathcal{U}^{\star}_{q}(K^{1-q}s_{0}(p))$, this implies immediately
that the convergence rate attained by the adaptive thresholding
estimator over $\mathcal{U}^{\star}_{q}$ in Theorem \ref{bpsth4}  and
(\ref{rate.expectation}) is optimal. }
\end{remark}

\begin{remark}{\rm
The estimator $\hat {\boldsymbol{\Sigma}}^\star(\delta)$ yields immediately an estimate of  the correlation matrix $\boldsymbol{R}_0=(r_{ij})_{1\le i, j, \le p}$ which is the object of direct interest in some statistical applications. Denote the corresponding estimator of $\boldsymbol{R}_0$ by $\hat{\boldsymbol R}^\star(\delta)=(\hat r^\star_{ij})_{1\le i, j, \le p}$ with $\hat r^\star_{ij} = \hat \sigma^\star_{ij}/\sqrt{\hat \sigma_{ii}\hat \sigma_{jj}}$.
A parameter space for the correlation matrices is the following $\ell_{q}$ ball:
\beq
\label{corr.space}
\mathcal{R}^{\star}_{q}:=\mathcal{R}^{\star}_{q}(s_{0}(p))=\Big{\{}\boldsymbol{R}: \boldsymbol{R}\succ 0, ~
\max_{i}\sum_{j=1}^{p}|r_{ij}|^{q}\le s_{0}(p)\Big{\}}.
\eeq
Then Theorem \ref{bpsth4} holds for estimating the correlation matrix $\boldsymbol{R}_0$ by replacing $\hat{\boldsymbol{\Sigma}}^\star(\delta)$, $\boldsymbol{\Sigma}_0$ and $\mathcal{U}^{\star}_{q}$ with
$\hat{\boldsymbol{R}}^\star(\delta)$, $\boldsymbol{R}_0$ and $\mathcal{R}^{\star}_{q}$, respectively.

Note that the covariance matrix $\boldsymbol{\Sigma}_0$ can be written as $\boldsymbol{\Sigma}_0=\boldsymbol{D}^{1/2}\boldsymbol{R}_0 \boldsymbol{D}^{1/2}$, where  $\boldsymbol{D}={\rm diag}(\boldsymbol{\Sigma}_{0})$. The covariance matrix can thus be viewed as a weighted version of the correlation matrix with weights $\{(\sigma_{ii}\sigma_{jj})^{1/2}\}$. Correspondingly, the parameter space $\mathcal{U}^{\star}_{q}$ in (\ref{cov.space})
can be viewed as the  weighted version of $\mathcal{R}^{\star}_{q}$ given in (\ref{corr.space}) with the same weights,
\[
\mathcal{U}^{\star}_{q}:=\Big{\{}\boldsymbol{\Sigma}: \boldsymbol{\Sigma}\succ 0, ~
\max_{i}\sum_{j=1}^{p}(\sigma_{ii}\sigma_{jj})^{1/2}|r_{ij}|^{q}\le s_{0}(p)\Big{\}}.
\]

}
\end{remark}

\subsection{Support recovery}

A closely related problem to estimating a sparse covariance matrix
under spectral norm is the recovery of the support of the covariance
matrix. This problem has been considered, for example, in Rothman,
Levina and Zhu (2009). For support recovery, it is natural to
consider the parameter space
\begin{eqnarray*}
\bar{\mathcal{U}}_{0} :=\bar{\mathcal{U}}_{0}(s_0(p)) =
\Big{\{}\boldsymbol{\Sigma}: \max_{i}\sum_{j=1}^{p}I\{\sigma_{ij}\neq 0\}\leq
s_{0}(p)\Big{\}},
\end{eqnarray*}
which assumes that the covariance matrix has at most $s_0(p)$
nonzero entries on each row.

Define the support of $\boldsymbol{\Sigma}_0=(\sigma_{ij}^0)$
by $\Psi=\{(i,j): \sigma^{0}_{ij}\neq 0\}$. The following theorem
shows that the adaptive thresholding estimator
$\hat{\boldsymbol{\Sigma}}^{\star}(\delta)$ recovers the support $\Psi$ exactly
with high probability when the magnitudes of nonzero entries are
above certain threshold.

\begin{theorem}\label{t2} Suppose that $\boldsymbol{\Sigma}_{0}\in\bar{\mathcal{U}}_{0}$.
Let $\delta\geq 2$ and
\begin{eqnarray}\label{t1}
|\sigma^{0}_{ij}|>(2+\delta+\gamma)\sqrt{\frac{\theta_{ij}\log p}{n}}\quad\mbox{for all $(i,j)\in\Psi$ and some $\gamma>0$.}
\end{eqnarray}
If either (C1) or (C2) holds, then we have
\begin{eqnarray*}
\inf_{\boldsymbol{\Sigma}_{0}\in \bar{\mathcal{U}}_{0}}\pr\Big{(}{\rm
supp}(\hat{\boldsymbol{\Sigma}}^{\star}(\delta))={\rm
supp}(\boldsymbol{\Sigma}_{0})\Big{)}\rightarrow 1.
\end{eqnarray*}
\end{theorem}
Similar support recovery result was established for the generalized universal
thresholding estimator in Rothman, Levina and Zhu (2009) under the
condition $\max_{i}\sigma^{0}_{ii}\leq K$ and a lower bound condition
similar to (\ref{t1}). Note that in Theorem \ref{t2}, we
do not require $\max_{i}\sigma_{ii}\leq K$.

Following Rothman, Levina and Zhu (2009), the ability to recover the support can be evaluated via the true positive rate (TPR) in combination with
the false positive rate (FPR), defined respectively as
\[
TPR=\frac{\#\{(i,j): \hat{\sigma}^{\star}_{ij}\neq 0\mbox{~and~}\sigma_{ij}\neq 0\}}{\#\{(i,j): \sigma_{ij}\neq 0\}} \; \mbox{ \rm and } \; FPR=\frac{\#\{(i,j):
\hat{\sigma}^{\star}_{ij}\neq 0\mbox{~and~}\sigma_{ij}=0\}}{\#\{(i,j): \sigma_{ij}= 0\}}.
\]
It follows from Theorem \ref{t2} directly that $\pr(FPR=0)\rightarrow 1$ and $\pr(TPR=1)\rightarrow 1$ under the conditions of the theorem.

The next result shows that $\delta=2$ is the optimal choice for
support recovery in the sense that a thresholding estimator with any
smaller choice of $\delta$ would fail to recover the support of
$\boldsymbol{\Sigma}_{0}$ exactly with probability going to one. We assume
$\textbf{X}$ satisfies the following condition which is weaker than
the Gaussian assumption.

 {\bf (C3)} Suppose that
$$\ep[ (X_{i}-\mu_{i})^{2}(X_{j}-\mu_{j})(X_{k}-\mu_{k})]=0,\quad
\ep[
(X_{i}-\mu_{i})(X_{j}-\mu_{j})(X_{k}-\mu_{k})(X_{l}-\mu_{l})]=0$$ if
$\sigma^{0}_{j_{1}j_{2}}=0$ for all $j_{1}\neq j_{2}\in
\{i,j,k,l\}$.

\begin{theorem}\label{th4}  Let $\lambda_{ij}=\tau\sqrt{\frac{\hat{\theta}_{ij}\log
 p}{n}}$ with $0<\tau<2$. Suppose that (C1) or (C2) holds. Under (C3) and $p=\exp(o(n^{1/5}))$, if $s_{0}(p)=O(p^{1-\tau_{1}})$ with some $\tau^{2}/4<\tau_{1}<1$
 and $p\rightarrow\infty$, then
\begin{eqnarray*}
\inf_{\boldsymbol{\Sigma}_{0}\in \bar{\mathcal{U}}_{0}}\pr\Big{(}{\rm
supp}(\hat{\boldsymbol{\Sigma}}^{\star}(\tau))\neq {\rm
supp}(\boldsymbol{\Sigma}_{0})\Big{)}\rightarrow 1.
\end{eqnarray*}
\end{theorem}

\begin{remark}
{\rm    The condition $p=\exp(o(n^{1/5}))$ is used in the proof to
make sure the covariances of the samples $\{\textbf{X}_{n}\}$ can be
well approximated by normal vectors.
 It can be replaced by $p=\exp(o(n^{1/3}))$ if
$\textbf{X}$ is a multivariate normal population. }
\end{remark}

\subsection{Comparison with universal thresholding}

It is interesting to compare the asymptotic results for adaptive thresholding estimator $\hat{\boldsymbol{\Sigma}}^{\star}(\delta)$ with the known results for universal thresholding estimators.
We begin by comparing the rate of convergence of $\hat{\boldsymbol{\Sigma}}^{\star}(\delta)$ with that of the universal thresholding estimator $\hat{\boldsymbol{\Sigma}}_{u}$ introduced in Bickel
and Levina (2008) in the case of polynomial-type tails. Suppose that
(C2) holds. Bickel and Levina (2008) showed that
\begin{eqnarray}\label{cb1}
\|\hat{\boldsymbol{\Sigma}}_{u}-\boldsymbol{\Sigma}_{0}\|_{2}=O_{\pr}\Big{(}s_{0}(p)\Big{(}\frac{p^{1/(1+\gamma+\epsilon/2)}}{n^{1/2}}\Big{)}^{1-q}\Big{)}
\end{eqnarray}
for $\boldsymbol{\Sigma}_{0}\in\mathcal{U}_{q}$. Clearly, the convergence rate given in Theorem
\ref{bpsth4} for the adaptive thresholding estimator is significantly faster than that in (\ref{cb1}).

We next compare the rates over the class $\mathcal{U}^{\star}_{q}$,
$0\leq q<1$. For brevity, we shall focus on  the Gaussian case
$\textbf{X}\sim N(\boldsymbol{\mu},\boldsymbol{\Sigma}_{0})$. The following theorem gives the
lower bound of the universal thresholding estimator.

\begin{theorem}\label{th5}
Assume that  $ n^{5q}\leq p\leq
\exp(o(n^{1/3}))$  and $8\leq s_{0}(p)< \min\{p^{1/4},4(n/\log
p)^{1/2}\}$. We have, as $p\rightarrow\infty$,
\begin{eqnarray}\label{ap1}
\inf_{\lambda_{n}}\sup_{\boldsymbol{\Sigma}_{0}\in\mathcal{U}^{\star}_{q}}\pr\Big{(}\|\hat{\boldsymbol{\Sigma}}_{g}-\boldsymbol{\Sigma}_{0}\|_{2}>\frac{3}{64}
s^{2-q}_{0}(p)\Big{(}\frac{\log
p}{n}\Big{)}^{(1-q)/2}\Big{)}\rightarrow 1
\end{eqnarray}
and hence for large $n$,
\begin{eqnarray}\label{ap3}
\inf_{\lambda_{n}}\sup_{\boldsymbol{\Sigma}_{0}\in\mathcal{U}^{\star}_{q}}\ep\|\hat{\boldsymbol{\Sigma}}_{g}-\boldsymbol{\Sigma}_{0}\|^{2}_{2}\geq
\frac{1}{512}s^{4-2q}_{0}(p)\Big{(}\frac{\log p}{n}\Big{)}^{1-q}.
\end{eqnarray}
\end{theorem}

The rate in (\ref{ap1}) is slower than the optimal rate
$s_{0}(p) (\log p/n)^{(1-q)/2}$ given in (\ref{thre-1}) when
$s_0(p)\goto \infty$ as $p\goto \infty$. Therefore
no universal thresholding estimators can be
 minimax-rate optimal under the spectral norm over
$\mathcal{U}^{\star}_{q}$ if $s_{0}(p)\rightarrow\infty$.


If we assume the mean of $\textbf{X}$ is zero and ignore the term
$\bar{\textbf{X}}$ in $\boldsymbol{\Sigma}_{n}$, then the universal thresholding estimators given in Bickel and Levina (2008) and Rothman, Levina and Zhu (2009) use the sample mean of the samples
$\{X_{ki}X_{kj};1\leq k\leq n\}$ to identify zero entries in the
covariance matrix.  The support of these estimators depends on the
quantities $I\{|\hat{\sigma}_{ij}|\geq \lambda_{n}\}$. In the high
dimensional setting, the sample mean is usually unstable for
non-Gaussian distributions with heavier tails. Non-Gaussian data can
often arise from many practical applications such as in finance and
genomics. For our estimator, instead of the sample mean, we
use the Student $t$ statistic
$\hat{\sigma}_{ij}/\hat{\theta}^{1/2}_{ij}$ to distinguish zero and
nonzero entries. Our support recovery depends on the quantities
$I\{|\hat{\sigma}_{ij}|/\hat{\theta}^{1/2}_{ij}\geq 2\sqrt{\log
p/n}\}$ which are more stable than $I\{|\hat{\sigma}_{ij}|\geq
\lambda_{n}\}$, since $t$ statistic is much more stable than the
sample mean; see Shao (1999) for the theoretical justification.

\section{Data-driven choice of $\delta$}
\label{delta.sec}

Section \ref{theory.sec} analyzes the properties of the adaptive
thresholding estimator with a fixed value of $\delta$.
Alternatively, $\delta$ can be selected empirically through cross
validation (CV). In Bickel and Levina (2008) the
value of the universal thresholding level $\lambda_{n}$ is not fully
specified and the CV method was used to select $\lambda_{n}$
empirically. They obtained the convergence rate under the Frobenius norm for an estimator that is
based only on partial samples. Theoretical analysis on  the rate of convergence
under the spectral norm is still lacking. In this section, we first
briefly describe the CV method for choosing $\delta$ and then derive
the theoretical properties of the resulting estimator under the
spectral norm.

Divide the sample $\{\textbf{X}_{k};1\leq k\leq n\}$ into two
subsamples at random. Let $n_{1}$ and $n_{2}=n-n_{1}$ be the two sample
sizes for the random split satisfying $n_{1}\asymp n_{2}\asymp n$,
and let $\hat{\boldsymbol{\Sigma}}^{v}_{1}$, $\hat{\boldsymbol{\Sigma}}^{v}_{2}$ be the two
sample covariance matrices from the $v$th split, for $v=1,\ldots,H$,
where $H$ is a fixed integer. Let $\hat{\boldsymbol{\Sigma}}^{\star
v}_{1}(\delta)$ and $\hat{\boldsymbol{\Sigma}}^{\star v}_{2}(\delta)$ be defined
as in (\ref{adap}) from the $v$th split and
\begin{eqnarray*}
\hat{R}(\delta)=\frac{1}{H}\sum_{v=1}^{H}\|\hat{\boldsymbol{\Sigma}}^{\star
v}_{1}(\delta)-\hat{\boldsymbol{\Sigma}}^{v}_{2}\|^{2}_{F}.
\end{eqnarray*}
Let $a_{j}=j/N$, $0\leq j\leq 4N$ be $4N+1$ points in $[0,4]$ and
take
\begin{eqnarray*}
\hat{\delta}=\hat{j}/N,\quad\mbox{where\quad}\hat{j}=\argmin_{0\leq
j\leq 4N}\hat{R}(j/N),
\end{eqnarray*}
where $N>0$ is a fixed integer.  If there are several $j$
attain the minimum value, $\hat{j}$ is chosen to be the smallest one. The final estimator of the covariance matrix
$\boldsymbol{\Sigma}_{0}$ is given by $\hat{\boldsymbol{\Sigma}}^{\star}(\hat \delta)$.

\begin{theorem}\label{th6}
Suppose $\textbf{X}\sim N(\boldsymbol{\mu},\boldsymbol{\Sigma}_{0})$ with $\boldsymbol{\Sigma}_{0}\in\mathcal{U}_{0}$ and $\min_{i}\sigma^{0}_{ii}\geq \tau_{0}$ for some
$\tau_{0}>0$. Let $s_{0}(p)=O((\log p)^{\gamma})$ for some
$\gamma<1$ and $n^{\xi}\leq p\leq \exp(o(n^{1/3}))$ for some
$\xi>0$. We have
\begin{eqnarray*}
\inf_{\boldsymbol{\Sigma}_{0}\in\mathcal{U}_{0}}\pr\Big{(}\|\hat{\boldsymbol{\Sigma}}^{\star}(\hat{\delta})-\boldsymbol{\Sigma}_{0}\|_{2}\leq
Cs_{0}(p)\Big{(}\frac{\log p}{n}\Big{)}^{1/2}\Big{)}\rightarrow 1.
\end{eqnarray*}
\end{theorem}

\begin{remark}
{\rm The assumption that $N$ is fixed  is not a stringent condition since
we only consider $\delta$ belonging to the fixed interval $[0,4]$.
 Moreover,
we will only focus on the matrices in $\mathcal{U}_{0}$ due to the
complexity of the proof. Extending to the case $N\rightarrow\infty$
with certain rate and more general $\boldsymbol{\Sigma}_{0}$ is possible.
However, it requires far more complicated proof and will not be
discussed in the present paper. }
\end{remark}

\begin{remark}
{\rm The condition $s_{0}(p)=O((\log p)^{\gamma})$ used in the theorem is purely for technical reasons and we believe that it is not essentially needed and can be weakened. This condition is not stringent when $p=\exp(n^{\alpha})$ and it becomes  restrictive if $p=O(n^{\alpha})$.
}
\end{remark}

Similar to the fixed $\delta$ case, we also consider support
recovery with the estimator $\hat{\boldsymbol{\Sigma}}^{\star}(\hat{\delta})$.
\bep\label{prop2}
 Suppose the conditions in Theorem \ref{th6} hold. For $\hat{\boldsymbol{\Sigma}}^{\star}(\hat{\delta})$, we have
\[
FPR=O_{\pr}(s_{0}(p)/p)\rightarrow 0.
\]
Moreover, since $\hat{\delta}\leq 4$, we have $TPR=1$ with probability tending to one if the lower bound in (\ref{t1}) holds with $2+\delta$ being
replaced by $6$. \eep


\section{Numerical Results}
\label{numerical.sec}

The adaptive thresholding procedure presented in Section \ref{method.sec} is easy to implement.
In this section, the numerical performance of the proposed adaptive thresholding estimator $\hat{\boldsymbol{\Sigma}}^{\star}(\delta)$ is studied using Monte Carlo simulations. Both methods for choosing the regularization parameter $\delta$ are considered and their performance are compared with that
of universal thresholding estimators. The adaptive thresholding estimator is illustrated in an analysis on a dataset from a small round blue-cell tumors  microarray experiment.

\subsection{Simulation}
\label{simu.sec}

The following two types of sparse covariance matrices are considered in the simulations to investigate the numerical properties of the adaptive thresholding estimator $\hat{\boldsymbol{\Sigma}}^{\star}(\delta)$ .

\begin{itemize}

 \item Model 1 (banded matrix with ordering). $\boldsymbol{\Sigma}_{0}={\rm diag}(\boldsymbol{A}_{1},\boldsymbol{A}_{2})$, where $\boldsymbol{A}_{1}=(\sigma_{ij})_{1\leq i,j\leq p/2}$,
 $\sigma_{ij}=\Big{(}1-\frac{|i-j|}{10}\Big{)}_{+}$,
 $\boldsymbol{A}_{2}=4\boldsymbol{I}_{p/2\times p/2}$. $\boldsymbol{\Sigma}_{0}$ is a two-block diagonal matrix. $\boldsymbol{A}_{1}$ is
a banded and sparse covariance matrix. $\boldsymbol{A}_{2}$ is a diagonal matrix with  $4$ along the diagonal.

\item Model 2 (sparse matrix without ordering). $\boldsymbol{\Sigma}_{0}={\rm diag}(\boldsymbol{A}_{1},\boldsymbol{A}_{2})$, where $\boldsymbol{A}_{2}=4\boldsymbol{I}_{p/2\times p/2}$, $\boldsymbol{A}_{1}=\boldsymbol{B}+\epsilon
\boldsymbol{I}_{p/2\times p/2}$, $\boldsymbol{B}=(b_{ij})_{p/2\times p/2}$ with independent
$b_{ij}={\rm unif}(0.3,0.8)\times {\rm Ber}(1,0.2)$. Here ${\rm
unif}(0.3,0.8)$ is a random variable taking value uniformly in
$[0.3, 0.8]$; ${\rm Ber}(1,0.2)$ is a Bernoulli random variable
which takes value 1 with probability 0.2 and  0 with
probability 0.8; and $\epsilon=\max(-\lambda_{\min}(\boldsymbol{B}),0)+0.01$ to
ensure that $\boldsymbol{A}_{1}$ is positive definite.


\end{itemize}

Under each model, $n=100$ independent and identically distributed
$p$-variate random vectors are generated from the normal
distribution with mean 0 and covariance matrix $\boldsymbol{\Sigma}_{0}$, for
$p=30,100,200$. In each setting, 100 replications are used. We
compare the numerical performance between the adaptive thresholding
estimators $\hat{\boldsymbol{\Sigma}}^{\star}(\hat{\delta})$ and
$\hat{\boldsymbol{\Sigma}}^{\star}_{2}\equiv \hat{\boldsymbol{\Sigma}}^{\star}(2)$ and with
the universal thresholding estimator $\hat{\boldsymbol{\Sigma}}_g$ of Rothman,
Levina and Zhu (2009). Here $\hat{\delta}$ is selected by five fold
cross-validation in Section \ref{delta.sec},
$\hat{\boldsymbol{\Sigma}}^{\star}_{2}$ is the
adaptive thresholding estimator with fixed $\delta=2$. The
thresholding level $\lambda_{n}$ in $\hat{\boldsymbol{\Sigma}}_g$ is selected by
five fold cross-validation method used in Bickel and Levina (2008).
For each procedure, we consider two types of thresholding functions,
the hard thresholding and the adaptive lasso thresholding
$s_{\lambda}(z)=x(1-|\lambda/x|^{\eta})$ with $\eta=4$. The losses
are measured by three matrix norms: the spectral norm, the matrix
$\ell_1$ norm and the Frobenius norm. We report in Tables
\ref{tb:simu1} and \ref{tb:simu2} the means and standard errors of
these losses.  We also carried out simulations with the SCAD
thresholding function for both universal thresholding and adaptive
thresholding. The phenomenon is very similar. The SCAD adaptive
thresholding also outperforms the SCAD universal thresholding. For
reasons of space, the results are not reported here.

\begin{table}[hptb]\small\addtolength{\tabcolsep}{-4pt}
  \begin{center}
    \caption{Comparison of average matrix losses for Model 1
      over $100$ replications. The standard errors are given in the parentheses.}
    \begin{tabular}{|r r r r @{\hspace{2em}} rrr| }

      \hline
      & \multicolumn{3}{c}{Adaptive lasso}  &\multicolumn{3}{c|}{Hard} \\[3pt]
      \hline
      \multicolumn{1}{|c}{$p$}& \multicolumn{1}{c}{$\hat{\boldsymbol{\Sigma}}_g$}&
      \multicolumn{1}{c}{$\hat{\boldsymbol{\Sigma}}^{\star}(\hat{\delta})$} &\multicolumn{1}{c}{$\hat{\boldsymbol{\Sigma}}^{\star}_{2}$}&
      \multicolumn{1}{c}{$\hat{\boldsymbol{\Sigma}}_g$}& \multicolumn{1}{c}{$\hat{\boldsymbol{\Sigma}}^{\star}(\hat{\delta})$}
       & \multicolumn{1}{c|}{$\hat{\boldsymbol{\Sigma}}^{\star}_{2}$}  \\[4pt]
       \hline
  \multicolumn{7}{|c|}{Operator norm}\\[4pt]
      30 &   $3.53(0.13)$&  $1.72(0.05)$  & $2.39(0.07)$  &   $3.50(0.14)$&  $1.77(0.05)$  & $1.77(0.04)$    \\
      100 &  $7.94(0.11)$ & $2.72(0.05)$ & $4.68(0.06)$   &   $8.64(0.07)$ & $2.57(0.05)$ & $3.04(0.05)$    \\
      200 &   $8.95(0.004)$ & $3.23(0.05)$ &$5.70(0.05)$   &  $8.95(0.004)$ & $3.02(0.05)$  &$3.77(0.05)$     \\[4pt]
\multicolumn{7}{|c|}{Matrix $\ell_1$ norm}\\[4pt]
      30   &  $5.29(0.15)$&  $2.57(0.08)$  & $3.34(0.09)$   & $5.71(0.15)$&  $2.60(0.09)$  & $2.70(0.06)$       \\
      100 &     $9.03(0.05)$ &  $4.15(0.07)$  &$6.39(0.09)$   &$9.24(0.03)$ &  $4.17(0.07)$  &$4.87(0.09)$ \\
      200 &      $9.35(0.01)$ & $4.90(0.07)$ & $7.64(0.07)$  &    $9.35(0.01)$ & $4.89(0.07)$  & $5.97(0.09)$  \\[4pt]
\multicolumn{7}{|c|}{Frobenius norm}\\[4pt]
      30 &$5.97(0.10)$ & $3.15(0.05)$  & $3.68(0.05)$  &  $6.58(0.09)$ & $3.29(0.05)$  & $3.29(0.04)$ \\
      100 &   $15.93(0.12)$&   $6.57(0.05)$  &$8.92(0.06)$  &  $16.88(0.03)$&   $6.79(0.06)$  &$7.53(0.05)$\\
      200 &  $24.23(0.01)$ & $9.62(0.05)$  & $14.20(0.07)$ & $24.24(0.01)$ & $9.97(0.06)$  & $11.68(0.05)$\\
\hline
    \end{tabular}

    \label{tb:simu1}
  \end{center}
\end{table}

\begin{table}[hptb]\small\addtolength{\tabcolsep}{-4pt}
  \begin{center}
    \caption{Comparison of average matrix losses for Model 2
      over $100$ replications. The standard errors are given in the parentheses.}
    \begin{tabular}{|r r r  r@{\hspace{2em}} r r r| }
      \hline
      & \multicolumn{3}{c}{Adaptive lasso}  &\multicolumn{3}{c|}{Hard} \\[3pt]
      \hline
      \multicolumn{1}{|c}{$p$}& \multicolumn{1}{c}{$\hat{\boldsymbol{\Sigma}}_g$}&
      \multicolumn{1}{c}{$\hat{\boldsymbol{\Sigma}}^{\star}(\hat{\delta})$} &\multicolumn{1}{c}{$\hat{\boldsymbol{\Sigma}}^{\star}_{2}$}&
      \multicolumn{1}{c}{$\hat{\boldsymbol{\Sigma}}_g$}& \multicolumn{1}{c}{$\hat{\boldsymbol{\Sigma}}^{\star}(\hat{\delta})$}
       & \multicolumn{1}{c|}{$\hat{\boldsymbol{\Sigma}}^{\star}_{2}$}  \\[4pt]
       \hline
       \multicolumn{7}{|c|}{Operator norm}\\[4pt]
      30 &   $1.48(0.02)$&  $1.24(0.03)$  & $1.19(0.03)$  &   $1.50(0.02)$  & $1.25(0.03)$ &     $1.21(0.03)$   \\
      100 &  $5.31(0.01)$ & $2.82(0.05)$  & $4.71(0.03)$   &  $5.31(0.01)$ & $2.69(0.05)$  & $3.97(0.04)$        \\
      200 &   $10.74(0.01)$ & $6.78(0.08)$  &$10.52(0.02)$   & $10.74(0.01)$ & $6.58(0.10)$& $10.04(0.03)$         \\[4pt]
\multicolumn{7}{|c|}{Matrix $\ell_1$ norm}\\[4pt]
      30   &  $1.70(0.03)$&  $1.33(0.04)$  & $1.22(0.03)$   & $1.70(0.02)$ & $1.32(0.04)$  & $1.24(0.03)$       \\
      100 &     $6.16(0.01)$ &  $4.10(0.05)$ &$5.52(0.03)$   & $6.16(0.01)$ & $4.20(0.06)$  & $5.22(0.03)$ \\
      200 &      $12.70(0.01)$ & $9.81(0.08)$  & $12.31(0.04)$     &$12.70(0.01)$ & $10.06(0.08)$ & $12.06(0.04)$\\[4pt]
\multicolumn{7}{|c|}{Frobenius norm}\\
      30 &$4.08(0.03)$ & $2.52(0.04)$ & $2.57(0.04)$  & $4.10(0.03)$ & $2.50(0.04)$ & $2.45(0.04)$\\
      100 &   $12.77(0.01)$&   $7.57(0.05)$ &$10.96(0.04)$  & $12.78(0.02)$ & $8.07(0.06)$  & $10.00(0.05)$\\
      200 &  $25.51(0.01)$ & $16.94(0.07)$  & $24.67(0.03)$ & $25.52(0.01)$ & $18.69(0.07)$ & $24.05(0.03)$\\
\hline
    \end{tabular}

    \label{tb:simu2}
  \end{center}
\end{table}

\clearpage

Under Model 1 and Model 2, both adaptive thresholding estimators
$\hat{\boldsymbol{\Sigma}}^{\star}(\hat{\delta})$ and $\hat{\boldsymbol{\Sigma}}^{\star}_{2}$
uniformly outperform the universal thresholding rule
$\hat{\boldsymbol{\Sigma}}_g$ significantly, regardless which thresholding
function or which loss function is used. Between
$\hat{\boldsymbol{\Sigma}}^{\star}(\hat{\delta})$ and $\hat{\boldsymbol{\Sigma}}^{\star}_{2}$,
$\hat{\boldsymbol{\Sigma}}^{\star}(\hat{\delta})$ performs better than
$\hat{\boldsymbol{\Sigma}}^{\star}_{2}$ in general. Between the two thresholding
functions, the hard thresholding rule outperforms the adaptive lasso
thresholding rule for $\hat{\boldsymbol{\Sigma}}^{\star}_{2}$, while the
difference is not significant for
$\hat{\boldsymbol{\Sigma}}^{\star}(\hat{\delta})$.
For both models, the behaviors of hard and adaptive lasso universal thresholding rules are very similar. They both tend to ``over-threshold'' and
remove many nonzero off-diagonal entries of the covariance matrices.

\begin{figure}[htbp]
  \centering
  Model 1\\
  \subfloat[][Truth
  ]{\includegraphics[width=0.25\textwidth]{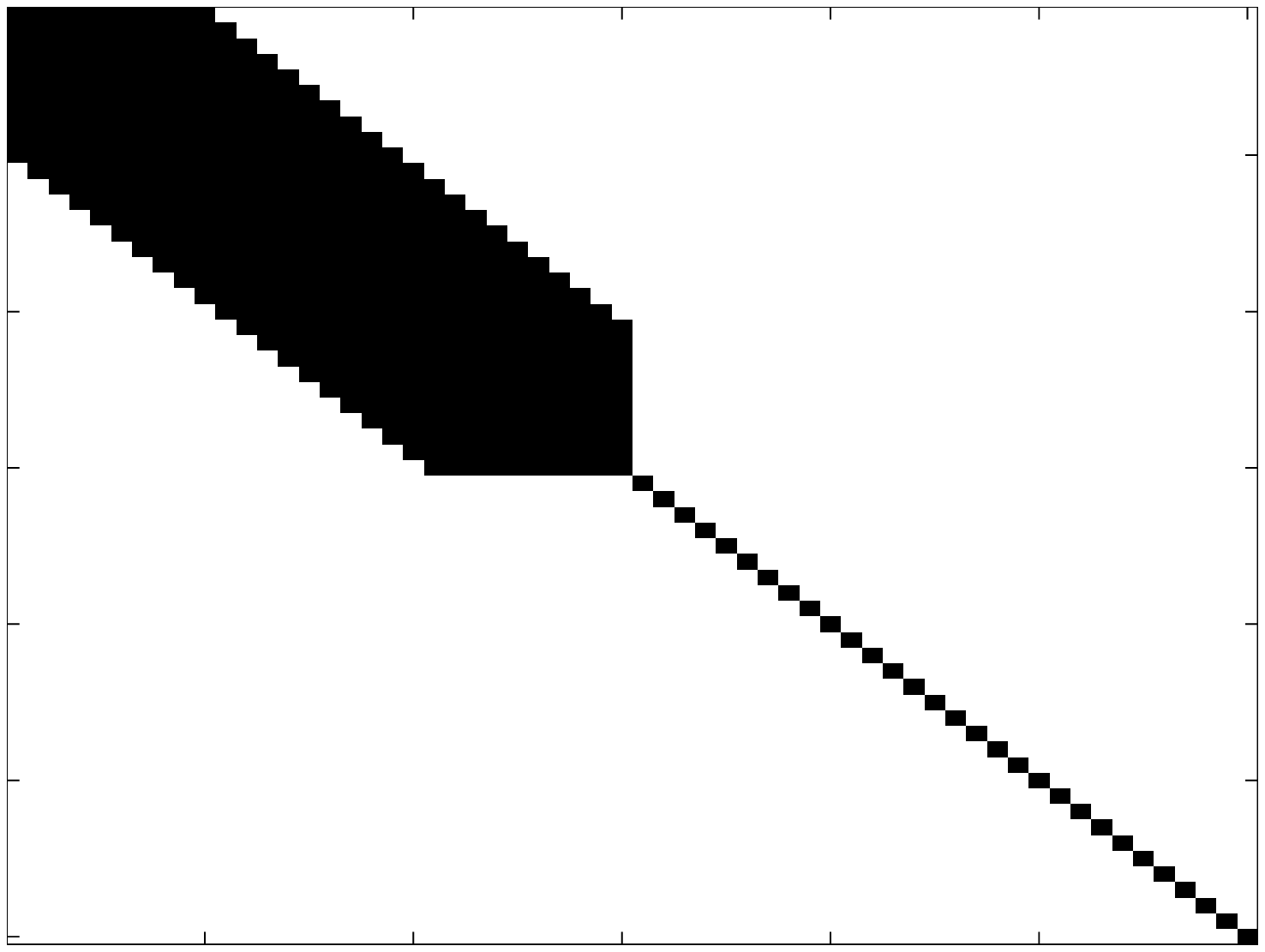} }
   \hspace*{0.1\textwidth}\subfloat[][$\hat{\boldsymbol{\Sigma}}(\hat{\delta})$(Hard)
  ]{\includegraphics[width=0.25\textwidth]{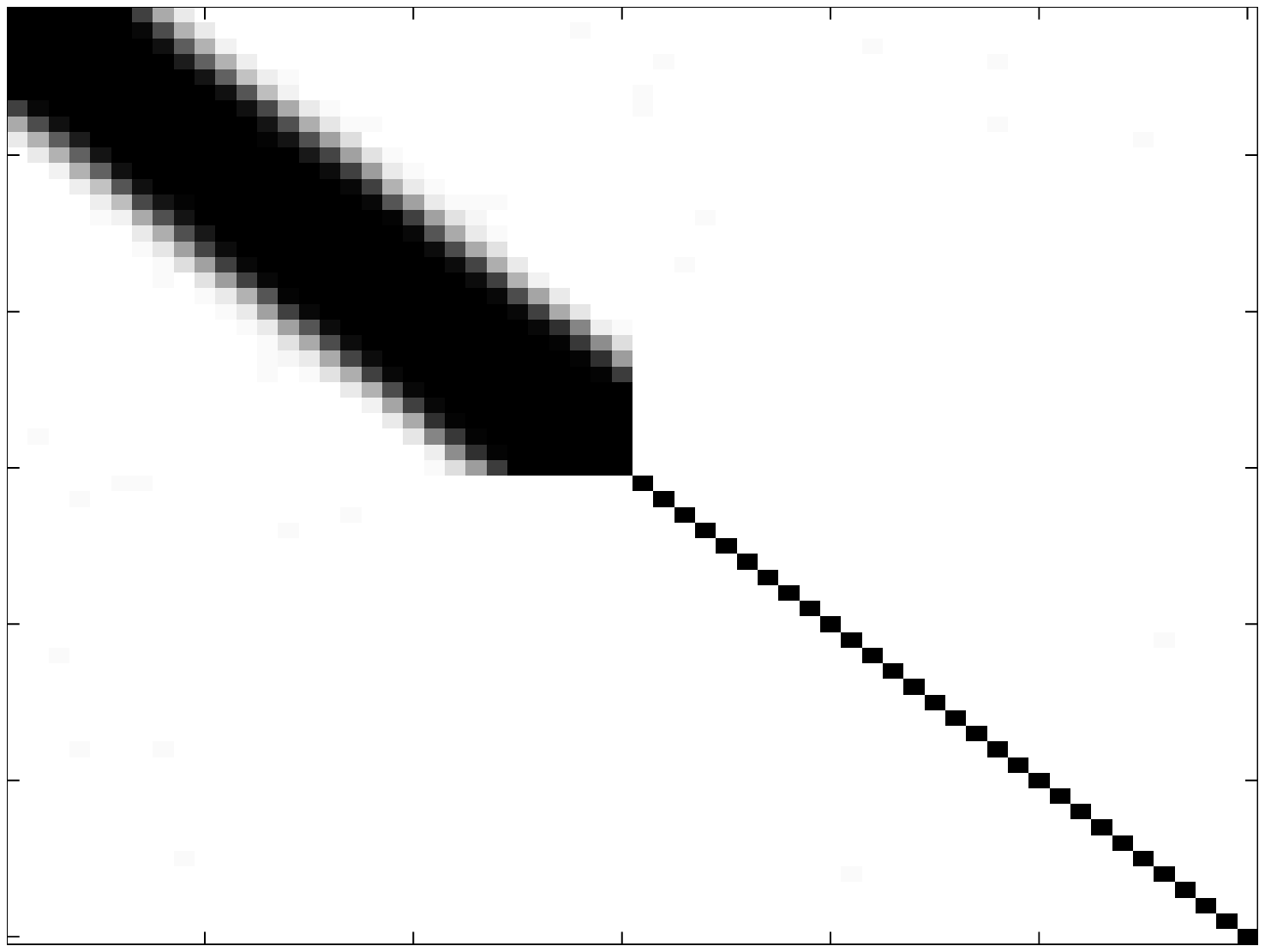} }
   \hspace*{0.1\textwidth} \subfloat[][$\hat{\boldsymbol{\Sigma}}(\hat{\delta})$(Adap.lasso)
  ]{\includegraphics[width=0.25\textwidth]{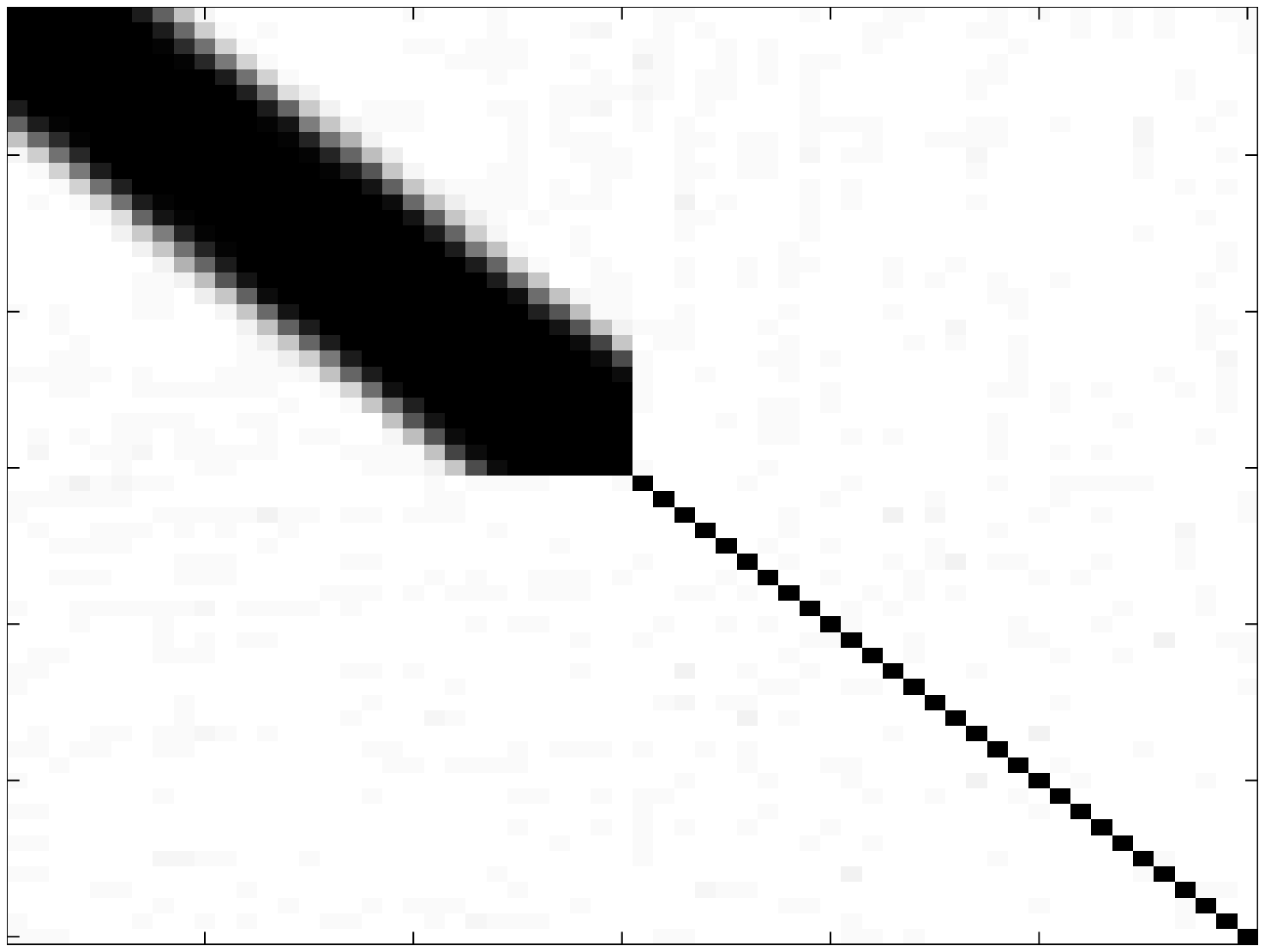} }
   \\
  \subfloat[][$\hat{\boldsymbol{\Sigma}}^{\star}_{2}$
  ]{\includegraphics[width=0.25\textwidth]{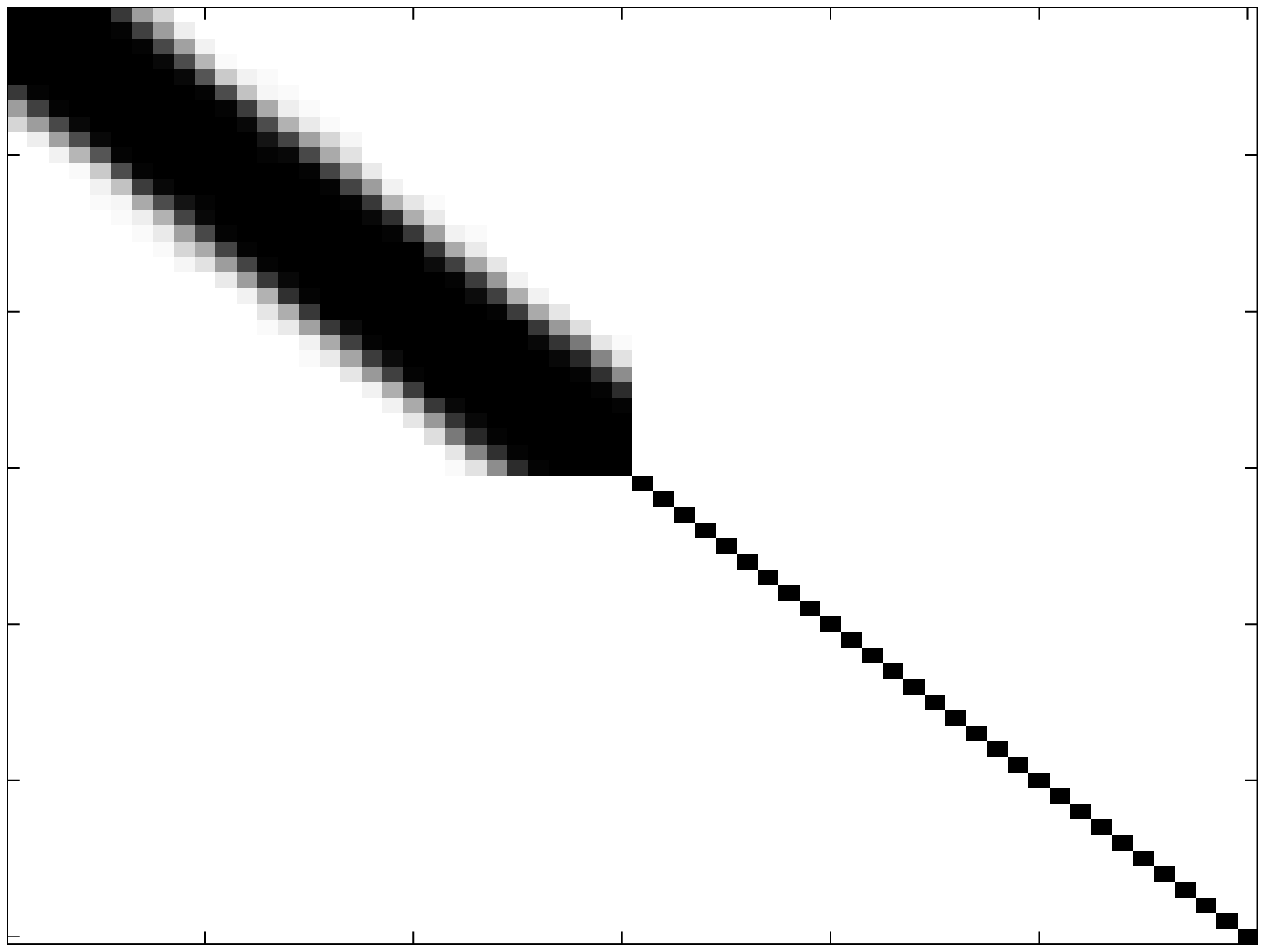} }
   \hspace*{0.1\textwidth}\subfloat[][$\hat{\boldsymbol{\Sigma}}_g$(Hard)
  ]{\includegraphics[width=0.25\textwidth]{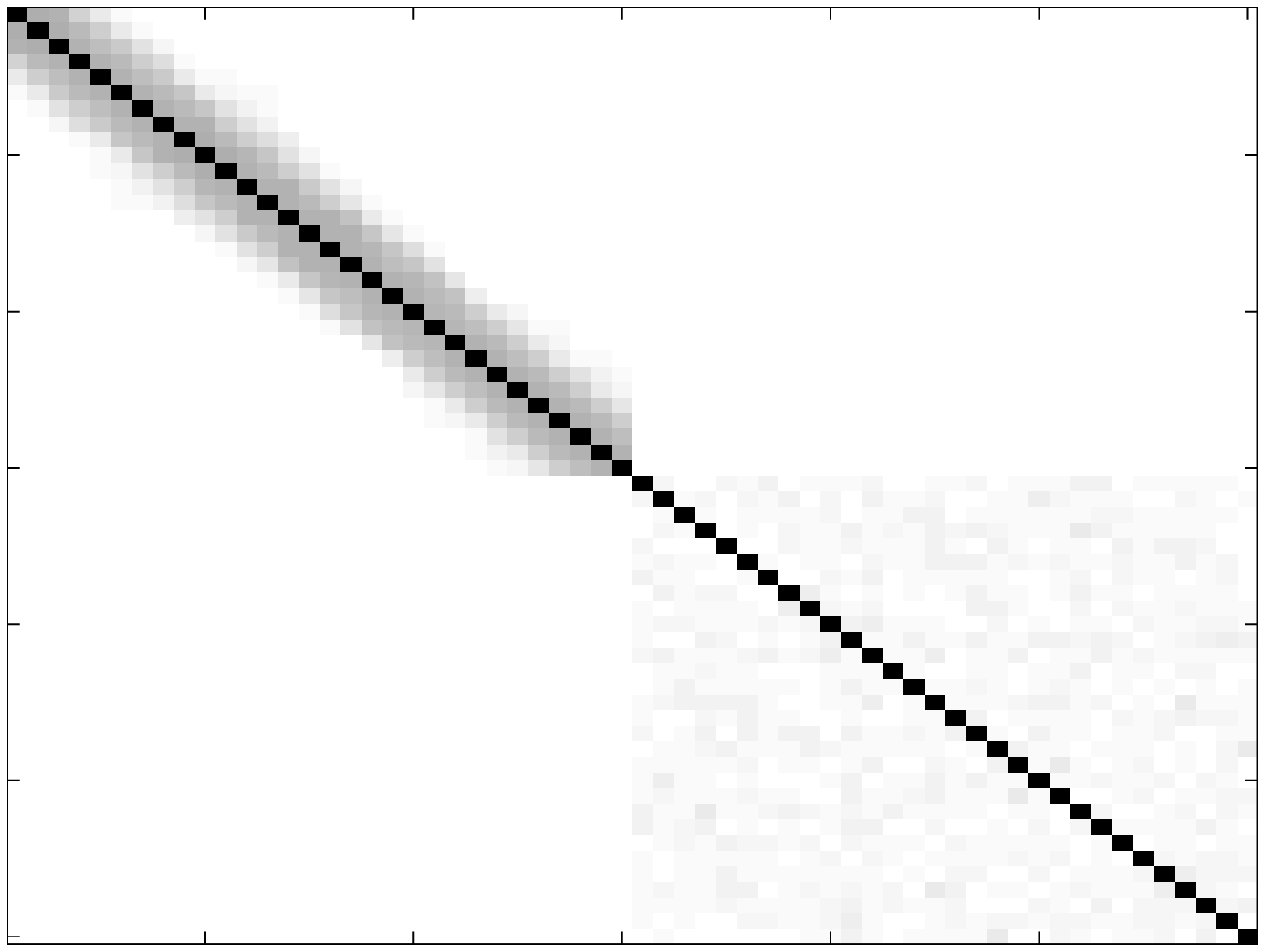} }
  \hspace*{0.1\textwidth}\subfloat[][$\hat{\boldsymbol{\Sigma}}_g$(Adap.lasso)
  ]{\includegraphics[width=0.25\textwidth]{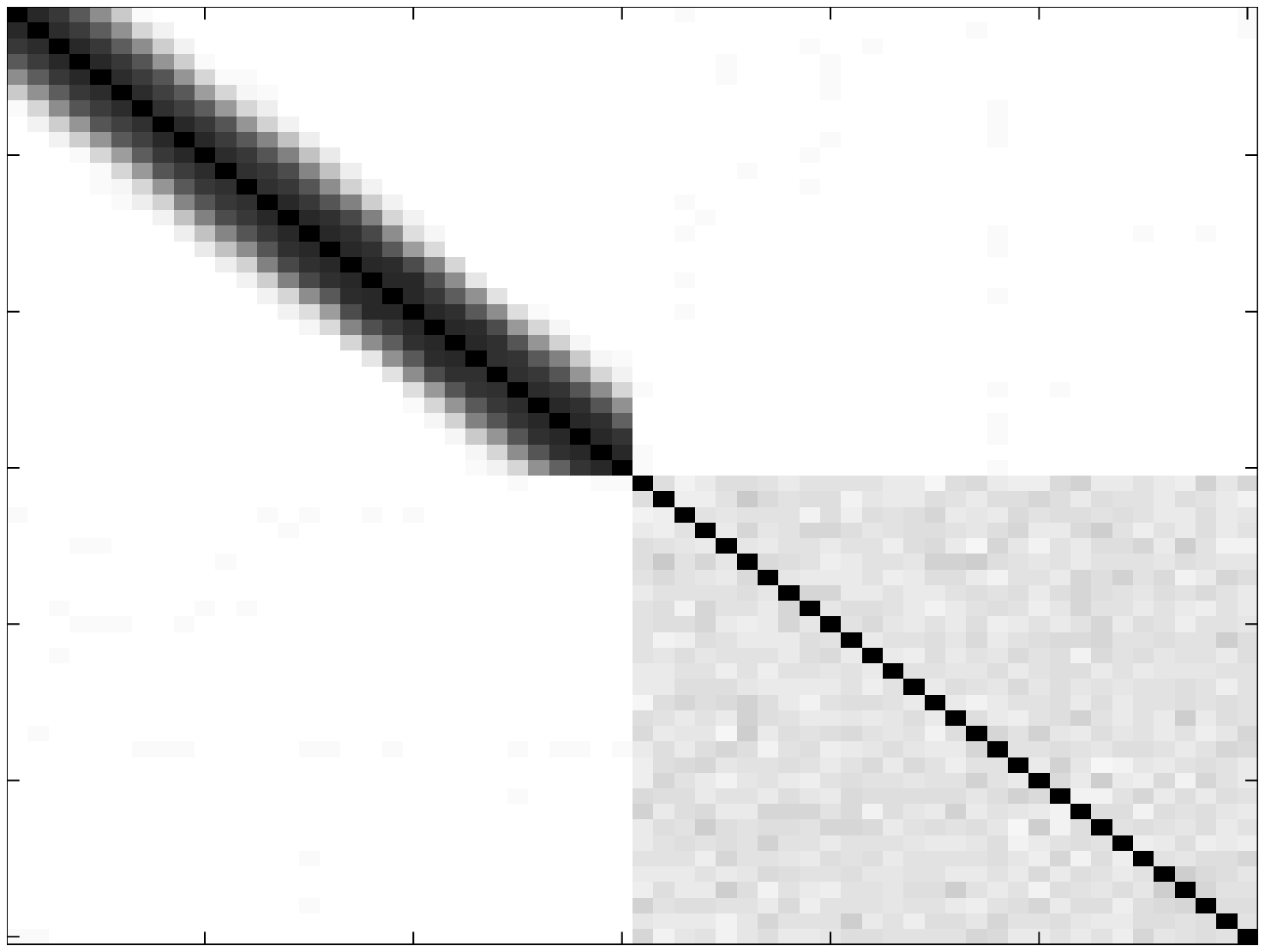} }
  \caption{Heat maps of the frequency of the zeros identified for each
    entry of the covariance matrix (when $p=60$) out of $100$
    replications.  White color is $100$ zeros identified out of $100$
    runs, and black is $0/100$.}
  \label{fig:heatmap1}
 \end{figure}
\begin{figure}[htbp]
  \centering
  Model 2\\
   \subfloat[][Truth
  ]{\includegraphics[width=0.25\textwidth]{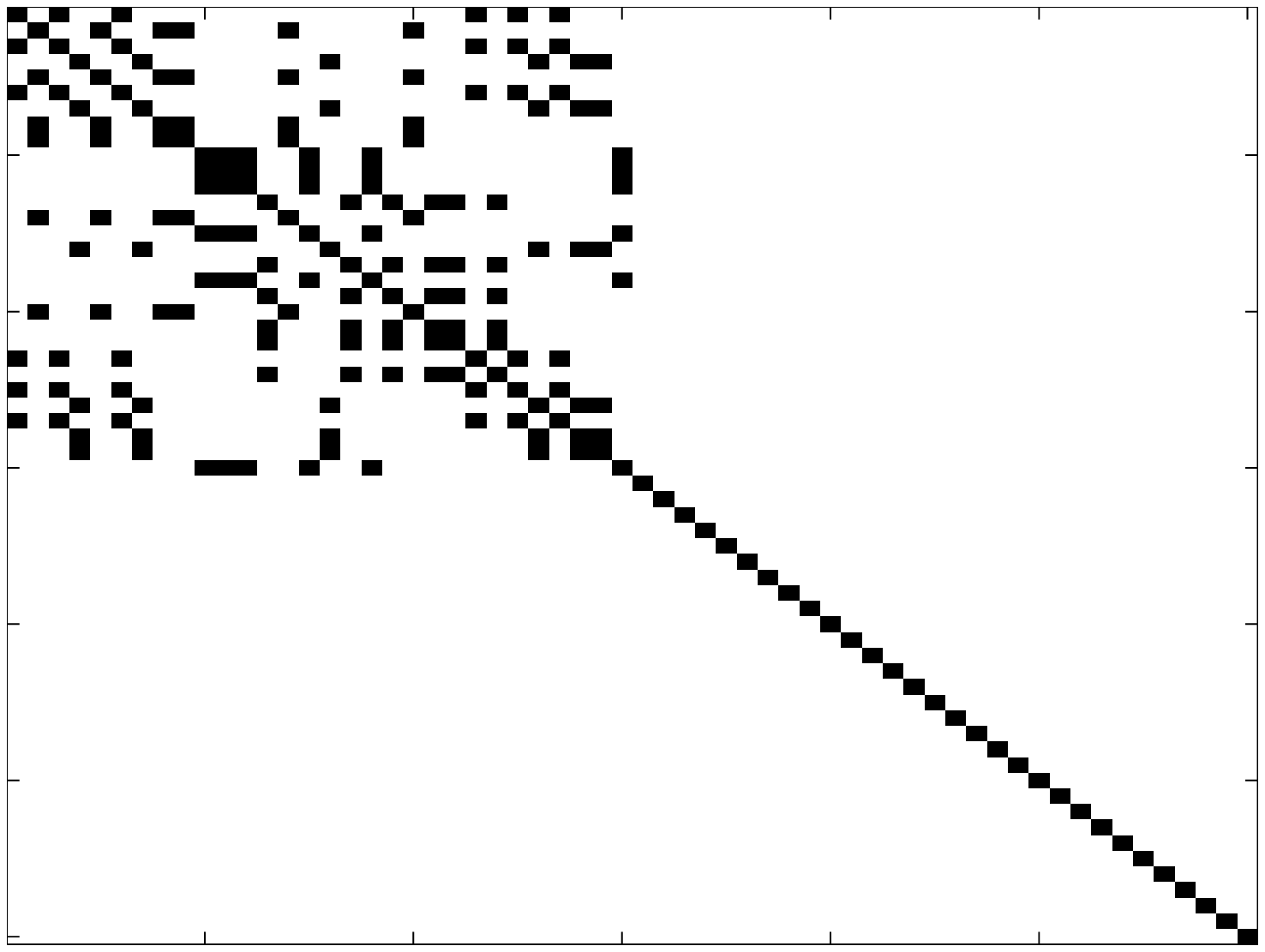} }
   \hspace*{0.1\textwidth}\subfloat[][$\hat{\boldsymbol{\Sigma}}(\hat{\delta})$(Hard)
  ]{\includegraphics[width=0.25\textwidth]{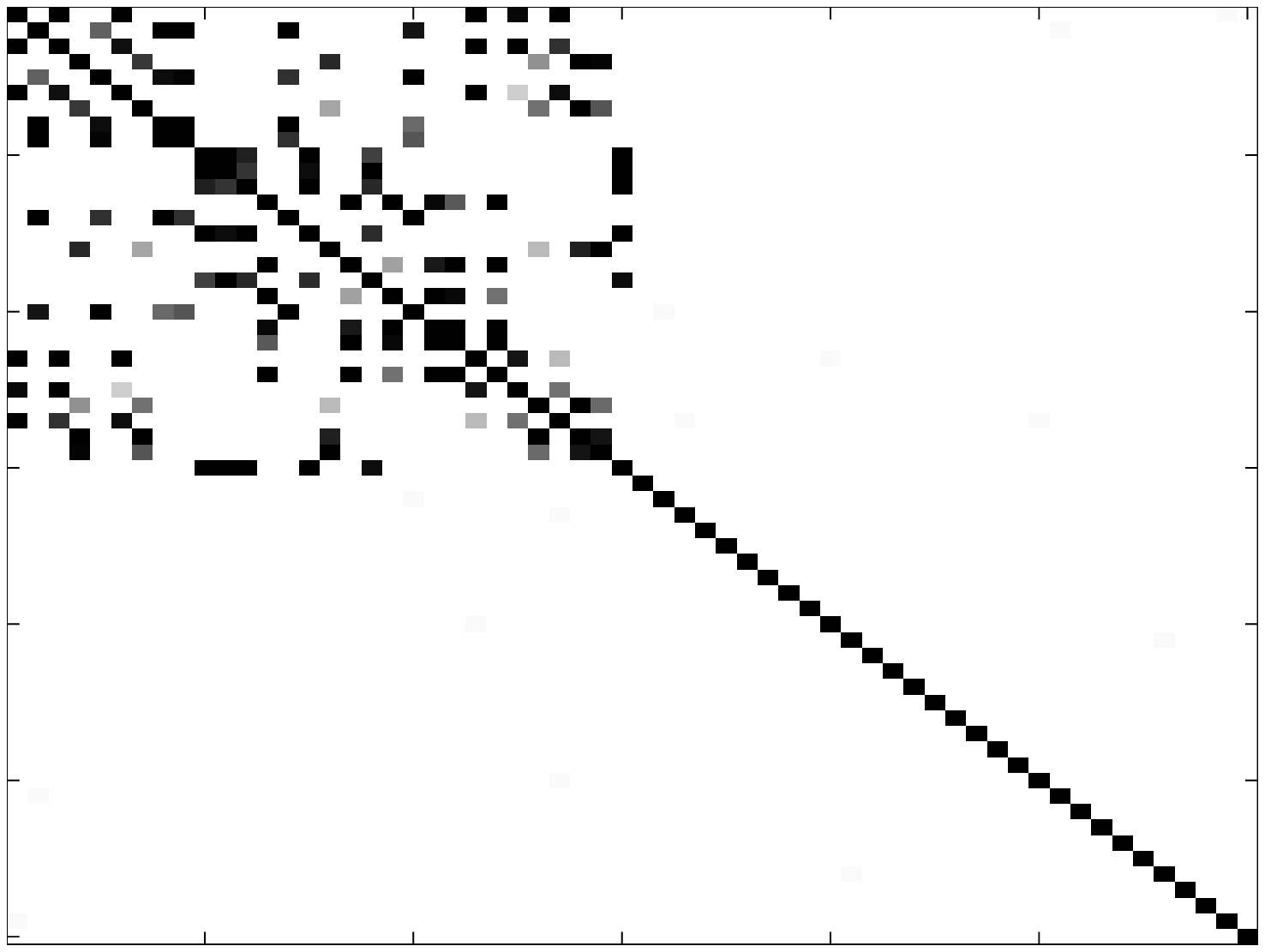} }
  \hspace*{0.1\textwidth}\subfloat[][$\hat{\boldsymbol{\Sigma}}(\hat{\delta})$(Adap.lasso)
  ]{\includegraphics[width=0.25\textwidth]{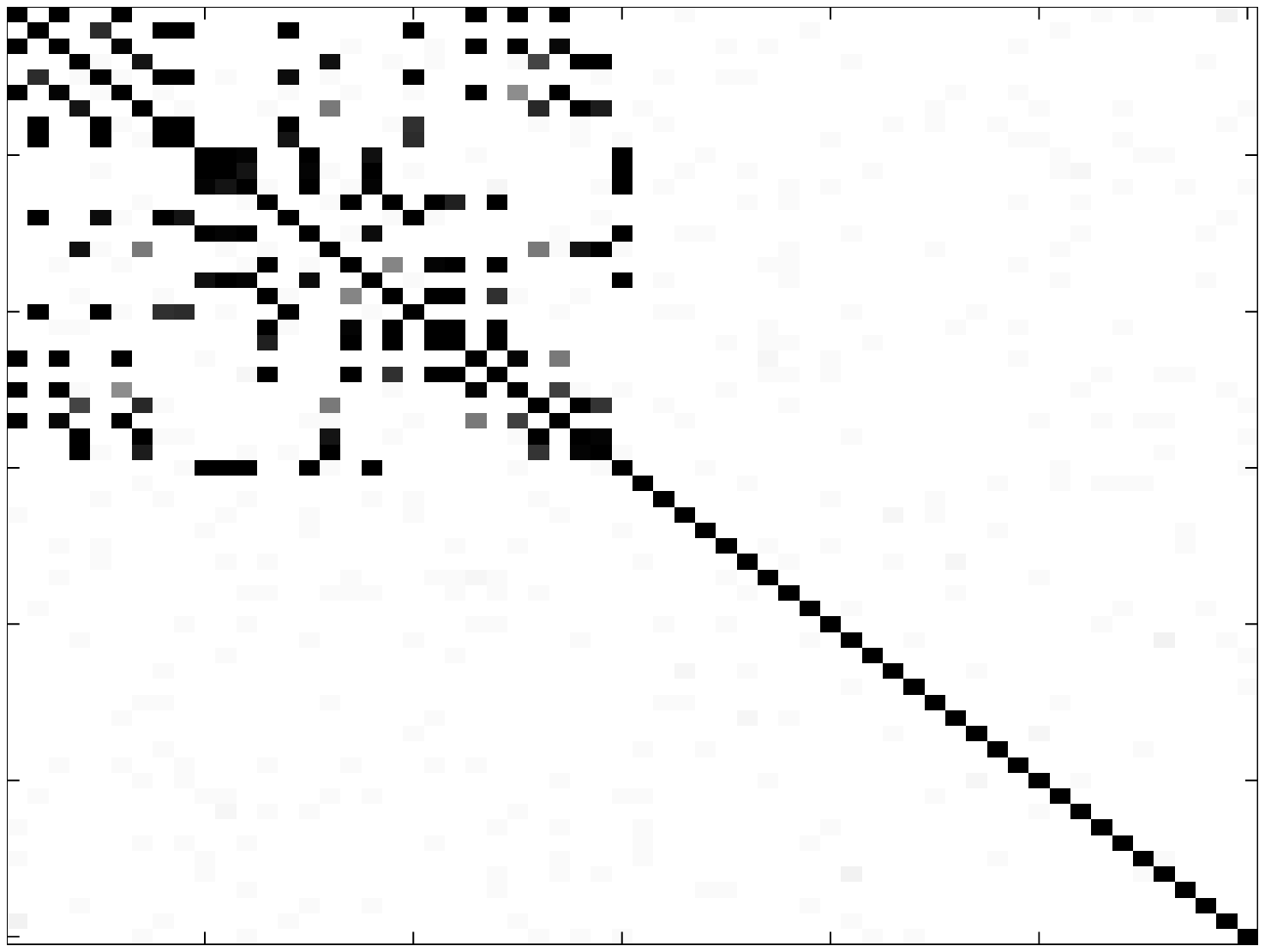} }
   \\
  \subfloat[][$\hat{\boldsymbol{\Sigma}}^{\star}_{2}$
  ]{\includegraphics[width=0.25\textwidth]{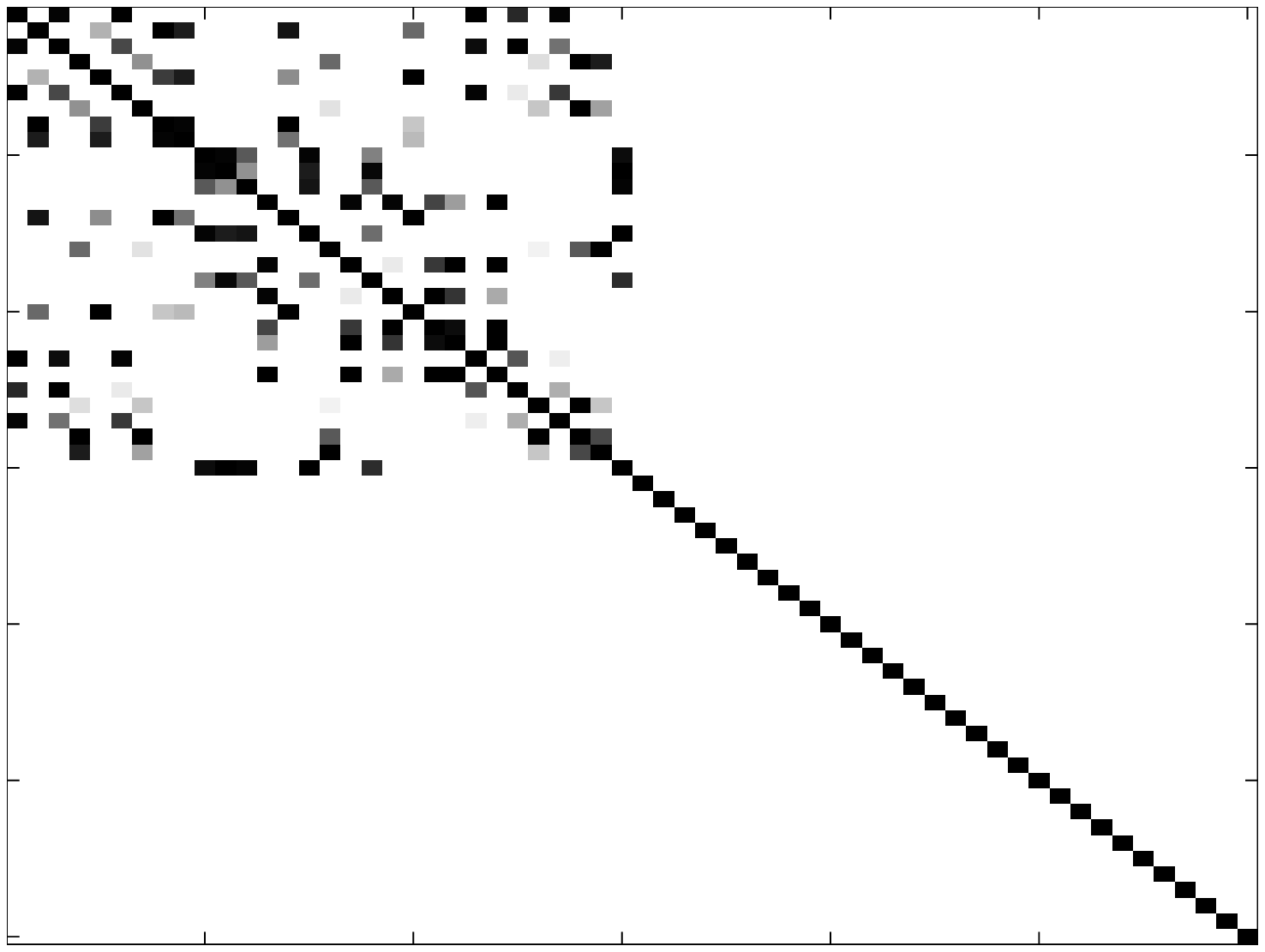} }
   \hspace*{0.1\textwidth}\subfloat[][$\hat{\boldsymbol{\Sigma}}_g$(Hard)
  ]{\includegraphics[width=0.25\textwidth]{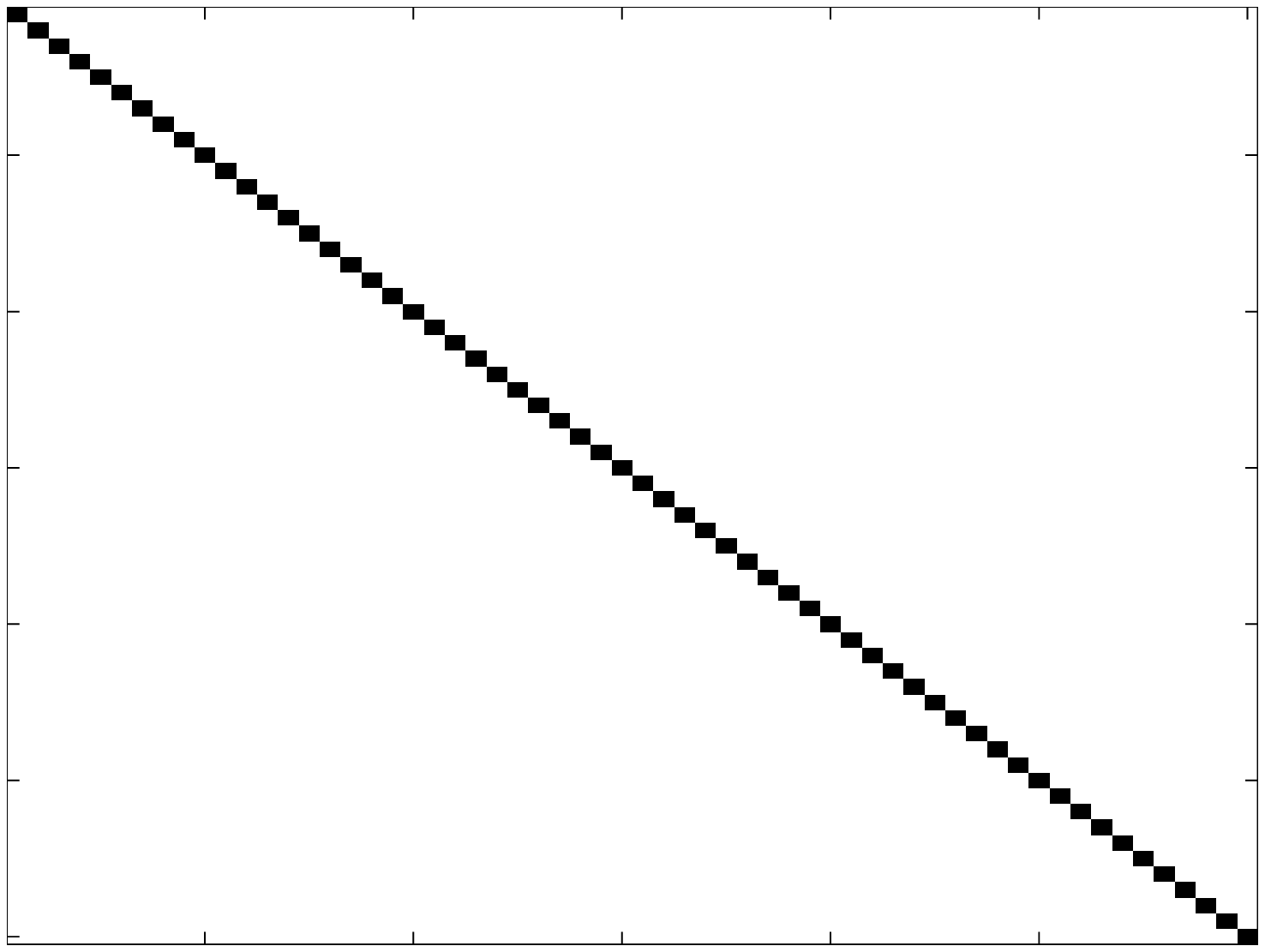} }
  \hspace*{0.1\textwidth}\subfloat[][$\hat{\boldsymbol{\Sigma}}_g$(Adap.lasso)
  ]{\includegraphics[width=0.25\textwidth]{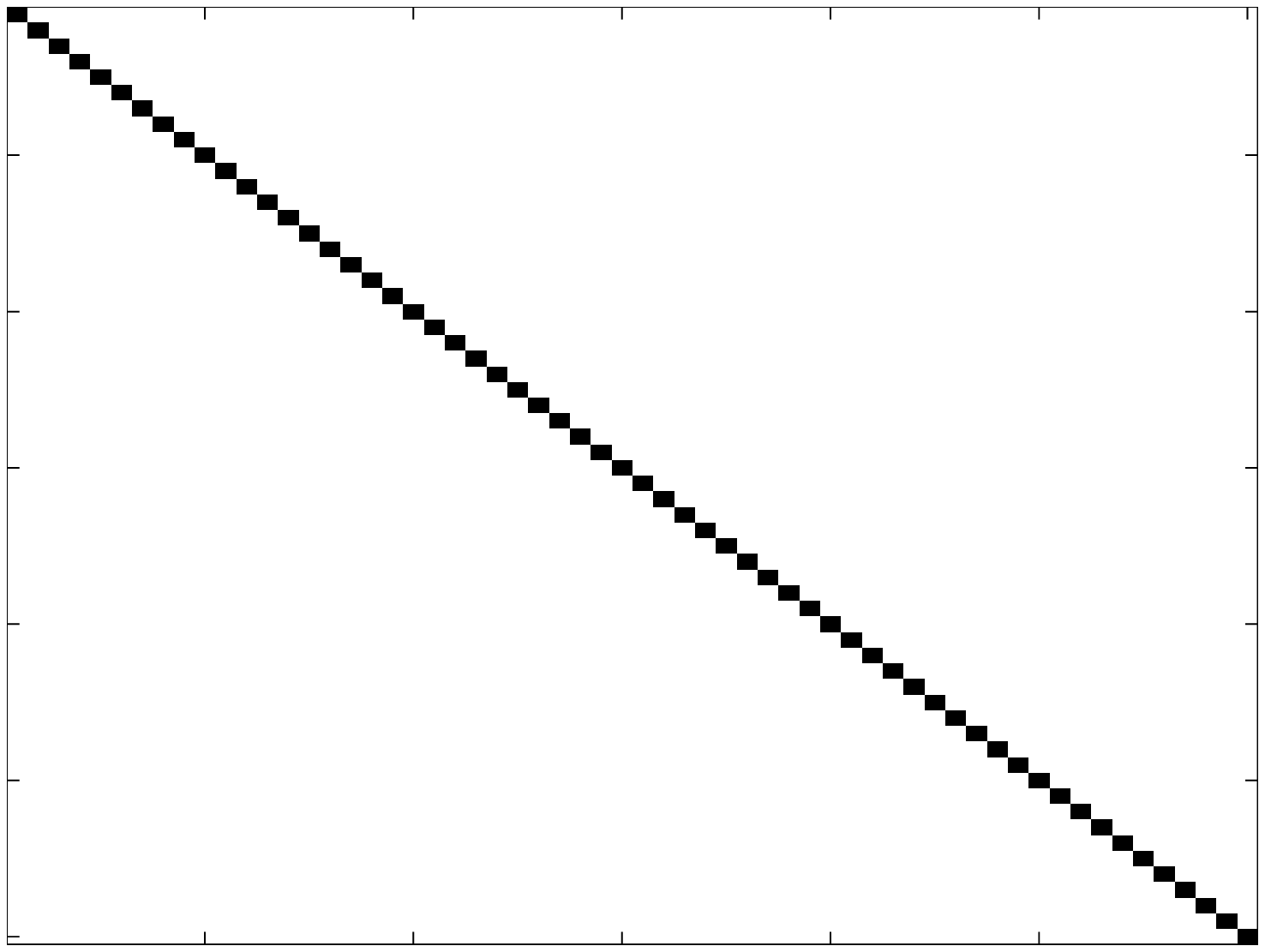} }
  \caption{Heat maps of the frequency of the zeros identified for each
    entry of the covariance matrix (when $p=60$) out of $100$
    replications.  White color is $100$ zeros identified out of $100$
    runs, and black is $0/100$.}
  \label{fig:heatmap2}
\end{figure}

For support recovery, again both
$\hat{\boldsymbol{\Sigma}}^{\star}(\hat{\delta})$ and $\hat{\boldsymbol{\Sigma}}^{\star}_{2}$
outperform $\hat{\boldsymbol{\Sigma}}_g$. The values of TPR and FPR based on the
off-diagonal entries are reported in Tables \ref{tb:simu3} and
\ref{tb:simu4}.  For Model 1, $\hat{\boldsymbol{\Sigma}}_{g}$ tends to estimate
many nonzero off-diagonal entries by zero when $p$ is large. To
better illustrate the recovery performance elementwise for the two
models, the heat maps of the nonzeros identified out of 100
replications when $p=60$ are pictured in Figures \ref{fig:heatmap1}
and \ref{fig:heatmap2}. The heat maps suggest that the sparsity
patterns recovered by $\hat{\boldsymbol{\Sigma}}^{\star}(\hat{\delta})$ and
$\hat{\boldsymbol{\Sigma}}^{\star}_{2}$ have significantly better resemblance to
the true model than $\hat{\boldsymbol{\Sigma}}_g$.


\begin{table}[hptb]\small\addtolength{\tabcolsep}{-4pt}
  \begin{center}
    \caption{Comparison of support recovery for Model 1 over 100 replications.}
    \begin{tabular}{|c c c c c@{\hspace{2em}} c c c| }
      \hline
      & \multicolumn{4}{c}{Adaptive lasso}  &\multicolumn{3}{c|}{Hard} \\
      \hline
      $p$& &$\hat{\boldsymbol{\Sigma}}_g$&  $\hat{\boldsymbol{\Sigma}}^{\star}(\hat{\delta})$ &$\hat{\boldsymbol{\Sigma}}^{\star}_{2}$&
      $\hat{\boldsymbol{\Sigma}}_g$& $\hat{\boldsymbol{\Sigma}}^{\star}(\hat{\delta})$    & $\hat{\boldsymbol{\Sigma}}^{\star}_{2}$  \\
      30 & TPR &  $0.57$&  $0.84$  & $0.72$  &  $0.46$&  $0.79$  & $0.72$     \\
         & FPR &  $0.07$&  $0.01$  & $0.00$  &  $0.05$&  $0.003$  & $0.00$     \\
      100 & TPR &   $0.15$ & $0.76$  & $0.57$   &  $0.008$ & $0.69$  & $0.57$        \\
          & FPR &   $0.01$ & $0.01$  & $0.00$   &  $0.00$ & $0.00$  & $0.00$        \\
      200 & TPR &   $0.00$ & $0.73$  &$0.51$   & $0.00$ & $0.65$  &$0.51$        \\
         & FPR &   $0.00$ & $0.003$  &$0.00$   & $0.00$ & $0.00$  &$0.00$        \\
\hline
    \end{tabular}

    \label{tb:simu3}
  \end{center}
\end{table}

\begin{table}[hptb]\small\addtolength{\tabcolsep}{-4pt}
  \begin{center}
    \caption{Comparison of support recovery for Model 2 over 100 replications.}
    \begin{tabular}{|c c c c c@{\hspace{2em}} c c c| }
      \hline
      & \multicolumn{4}{c}{Adaptive lasso} &\multicolumn{3}{c|}{Hard} \\
      \hline
      $p$& &$\hat{\boldsymbol{\Sigma}}_g$&  $\hat{\boldsymbol{\Sigma}}^{\star}(\hat{\delta})$  &$\hat{\boldsymbol{\Sigma}}^{\star}_{2}$&
      $\hat{\boldsymbol{\Sigma}}_g$& $\hat{\boldsymbol{\Sigma}}^{\star}(\hat{\delta})$   & $\hat{\boldsymbol{\Sigma}}^{\star}_{2}$  \\
      30 &   TPR & $0.02$&  $0.95$ & $0.88$  &   $0.00$  & $0.91$ &     $0.88$   \\
         & FPR &  $0.00$&  $0.01$ & $0.00$  &   $0.00$  & $0.00$ &     $0.00$   \\
      100 &  TPR & $0.00$ & $0.80$  & $0.33$   &  $0.00$ & $0.66$  & $0.33$        \\
          &  FPR & $0.00$ & $0.01$  & $0.00$   &  $0.00$ & $0.00$  & $0.00$        \\
      200 &  TPR & $0.00$ & $0.68$  &$0.09$   & $0.00$ & $0.49$ & $0.09$         \\
          &  FPR & $0.00$ & $0.01$  &$0.00$   & $0.00$ & $0.00$ & $0.00$         \\
\hline
    \end{tabular}

    \label{tb:simu4}
  \end{center}
\end{table}

\clearpage

\subsection{Correlation analysis on real data}
\label{realdata.sec}

We now apply  the adaptive thresholding estimator
$\hat{\boldsymbol{\Sigma}}^{\star}(\delta)$ to a dataset from a small round
blue-cell tumors (SRBC) microarray experiment (Khan et al., 2001)
and compare the ability of support recovery with that of the
universal thresholding estimator $\hat{\boldsymbol{\Sigma}}_g$. The estimator
$\hat{\boldsymbol{\Sigma}}^{\star}_{2}$ is not considered here since the
simulation results in Section \ref{simu.sec} show that
$\hat{\boldsymbol{\Sigma}}^{\star}(\hat{\delta})$ outperforms
$\hat{\boldsymbol{\Sigma}}^{\star}_{2}$ when the sample size is not large. The
SRBC data set has been analyzed in Rothman, Levina and Zhu (2009) in
which the universal thresholding rules were considered. To make the
results comparable, we shall follow the same steps as those in
Rothman, Levina and Zhu (2009).

The SRBC data has 63 training tissue samples, and 2308 gene expression values recorded for each sample. The original data has 6567 genes and was
reduced to 2308 genes  after an initial filtering; see Khan et al. (2001). The 63 tissue samples contain four types of tumors (23 EWS, 8 BL-NHL, 12 NB, and 20 RMS). As in Rothman, Levina and Zhu (2009),
the genes were first ranked by the amount of discriminative
information based on the F-statistic,
\begin{eqnarray*}
F=\frac{\frac{1}{k-1}\sum_{m=1}^{k}n_{m}(\bar{x}_{m}-\bar{x})^{2}}{\frac{1}{n-k}\sum_{m=1}^{k}(n_{m}-1)\hat{\sigma}^{2}_{m}},
\end{eqnarray*}
where $n=63$ is the sample size, $k=4$ is the number of classes,
$n_{m}$, $1\leq m\leq 4$, are the sample sizes of the four types of
tumors, $\bar{x}_{m}$ and $\hat{\sigma}_{m}$ are the sample mean and
sample variance of the class $m$, and $\bar{x}$ is the overall
sample mean. According to the $F$ values, the top 40 and bottom 160
genes were chosen. The first 40 genes were also ordered according to
the ordering given in Rothman, Levina and Zhu (2009). Based on the
200 genes, the performance of the two estimators
$\hat{\boldsymbol{\Sigma}}^{\star}(\hat{\delta})$ and $\hat{\boldsymbol{\Sigma}}_g$ was
considered.  The tuning parameters $\hat{\delta}$ and $\lambda_{n}$
were selected by five fold cross validation. To this end, we need to
divide the 63 samples into five groups of nearly equal sizes. As
there are four types of tumors in the samples, we let the
proportions of the four types of tumors in each group be nearly
equal so that each fold is a good representative of the whole. Three
fold cross validation is also used in this way and the results are
similar.

\begin{figure}[htbp]
\center
   \subfloat[][$\hat{\boldsymbol{\Sigma}}(\hat{\delta})$ Hard $(83.11\%$ zeros)
  ]{\includegraphics[width=0.35\textwidth]{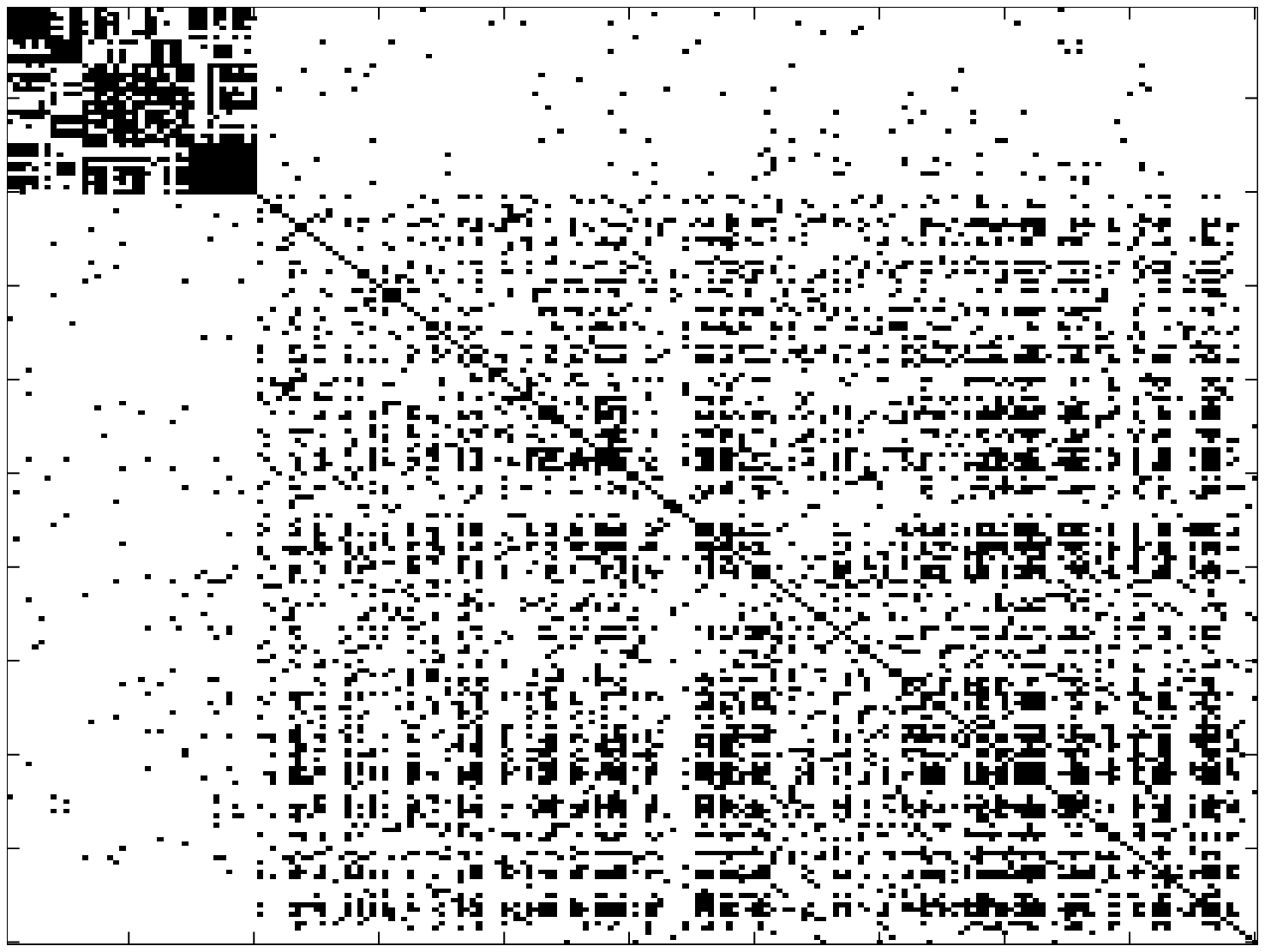} }
   \hspace*{0.1\textwidth} \subfloat[][$\hat{\boldsymbol{\Sigma}}(\hat{\delta})$ AL $(69.78\%$ zeros)
  ]{\includegraphics[width=0.35\textwidth]{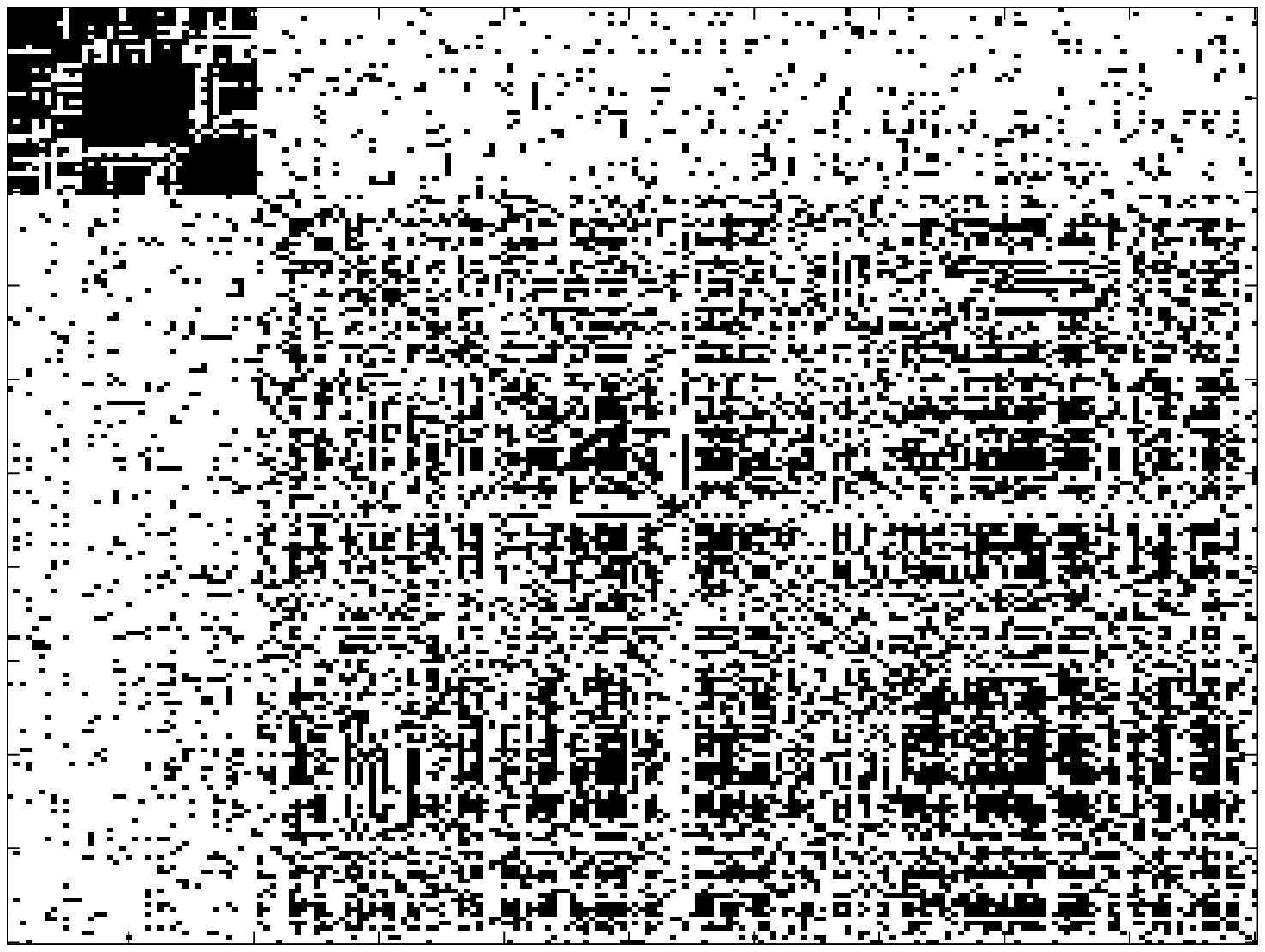} }\\
  \subfloat[][$\hat{\boldsymbol{\Sigma}}_g$ Hard $(97.88\%$ zeros)
  ]{\includegraphics[width=0.35\textwidth]{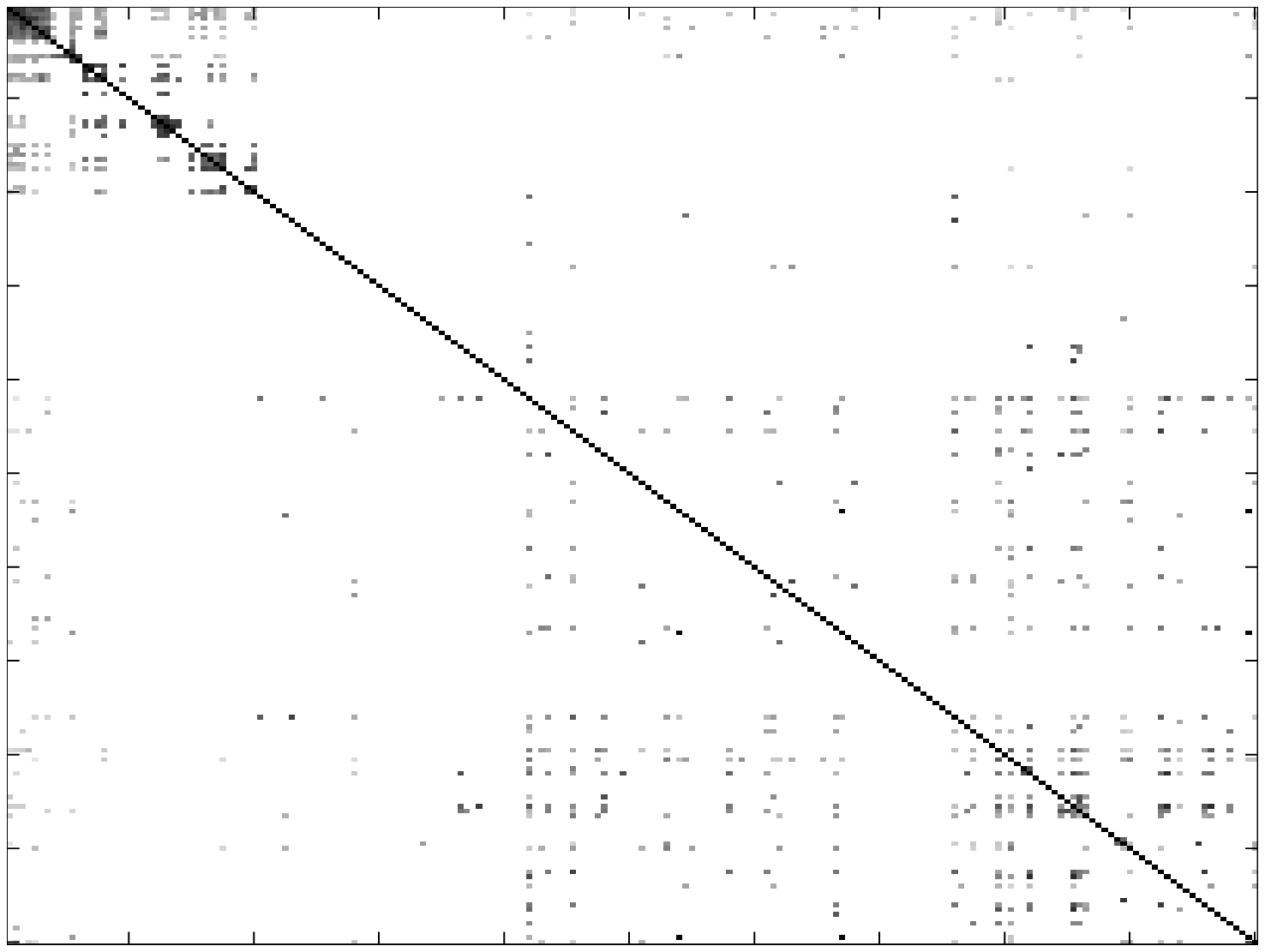} }
  \hspace*{0.1\textwidth}\subfloat[][$\hat{\boldsymbol{\Sigma}}_g$ AL $(97.88\%$ zeros)
  ]{\includegraphics[width=0.35\textwidth]{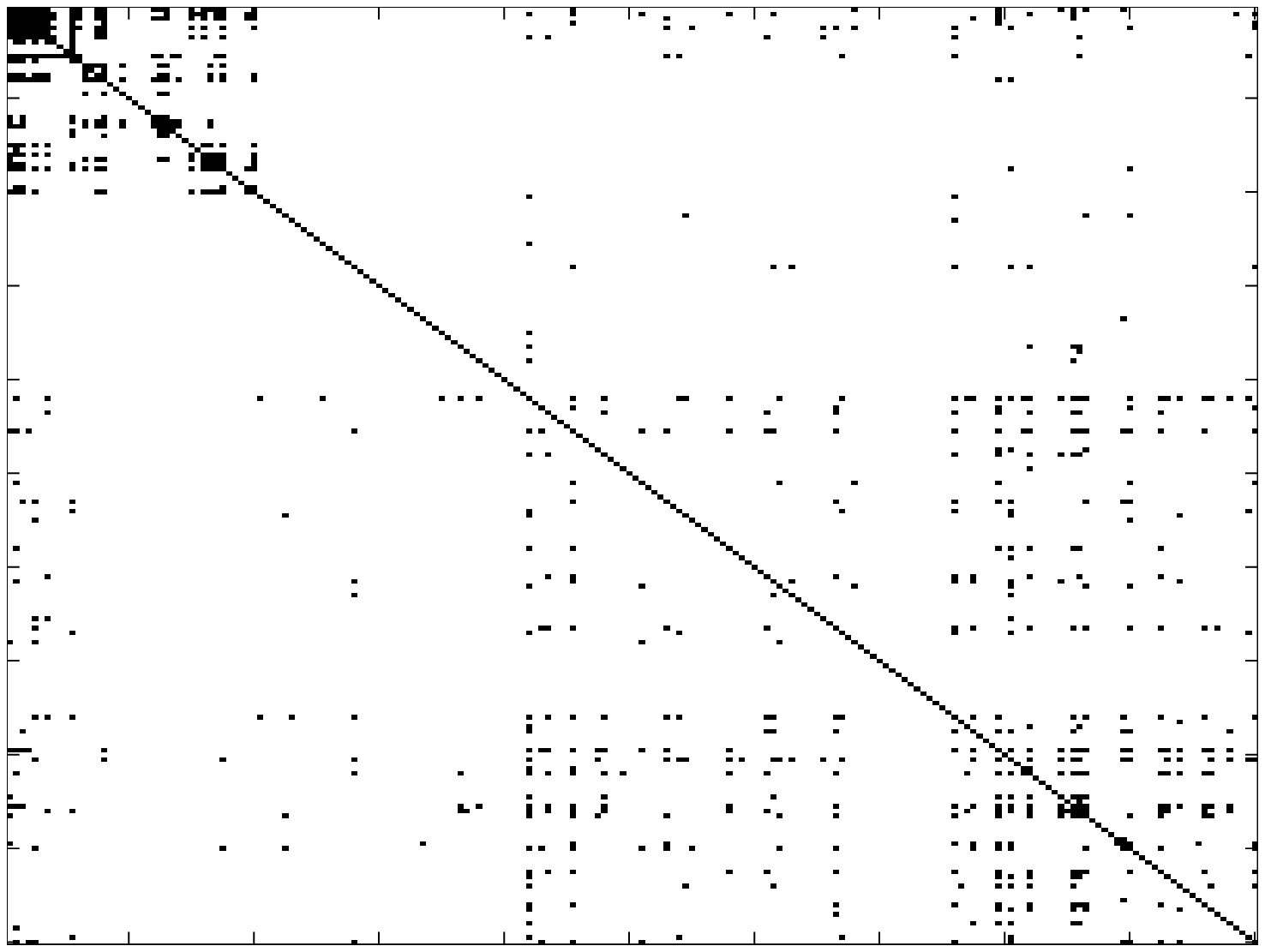} }
  \caption{Heatmaps of the estimated supports.  }
  \label{fig:heatmap3}
 \end{figure}

Figure \ref{fig:heatmap3} plots the heat maps of
$\hat{\boldsymbol{\Sigma}}^{\star}(\hat{\delta})$ with hard thresholding
($\hat{\boldsymbol{\Sigma}}^{\star}(\hat{\delta})$ Hard),  $\hat{\boldsymbol{\Sigma}}_g$ with
hard thresholding ($\hat{\boldsymbol{\Sigma}}_g$ Hard),
$\hat{\boldsymbol{\Sigma}}^{\star}(\hat{\delta})$ with adaptive lasso
thresholding ($\hat{\boldsymbol{\Sigma}}^{\star}(\hat{\delta})$ AL),
$\hat{\boldsymbol{\Sigma}}_g$ with adaptive lasso thresholding ($\hat{\boldsymbol{\Sigma}}_g$
AL).
$\hat{\boldsymbol{\Sigma}}_g$ AL and $\hat{\boldsymbol{\Sigma}}_g$ Hard result in very sparse
estimators, with $97.88\%$ zero elements in off diagonal positions.
The estimator $\hat{\boldsymbol{\Sigma}}^{\star}(\hat{\delta})$ AL
is the least sparse one with $69.78\%$ zeros, while $\hat{\boldsymbol{\Sigma}}^{\star}(\hat{\delta})$ Hard has $83.11\%$ zeros.
 The "over-threshold" phenomenon in the real data analysis is consistent with that observed in the simulations.
 The universal thresholding rule removes many nonzero off diagonal entries and results in an "over-sparse" estimate,
 while adaptive thresholding with different individual levels results in a clean but more informative
estimate of the sparsity structure.

\section{Discussion}
\label{discussion.sec}

This paper introduces an adaptive entry-dependent thresholding procedure for estimating sparse covariance matrices. The proposed estimator
$\hat{\boldsymbol{\Sigma}}^\star(\delta)=(\hat \sigma^\star_{ij})$ enjoys excellent performance both theoretically and numerically.  In particular, $\hat{\boldsymbol{\Sigma}}^\star(\delta)$ attains the optimal rate of convergence over $\mathcal{U}^{\star}_{q}$ given in (\ref{cov.space}) while universal thresholding estimators are shown to be sub-optimal.
The main reason that universal thresholding does not perform well is that the sample covariances can have a wide range of variabilities. A simple and natural way to deal with the heteroscedasticity is to first estimate the correlation matrix $\boldsymbol{R}_0$ and then renormalize by the sample variances to obtain an estimate of the covariance matrix. We shall discuss below two approaches based on this idea.

Denote the sample correlation matrix by  $\hat{\boldsymbol{R}} = (\hat r_{ij})_{1\le i, j\le p}$ with $\hat r_{i,j} = \hat \sigma_{ij}/\sqrt{\hat \sigma_{ii}\hat \sigma_{jj}}$. An estimate of the correlation matrix $\boldsymbol{R}_0$ can be obtained by thresholding $\hat r_{ij}$. Define the universal thresholding estimator of the correlation matrix by $\hat{\boldsymbol{R}}(\lambda_{n})=(\hat{r}^{thr}_{ij})_{p\times p}$ with
\[
\hat{r}^{thr}_{ij}=\hat{r}_{ij}I\{|\hat{r}_{ij}|\ge \lambda_{n}\}
\]
and the corresponding estimator of the covariance matrix by $\hat{\boldsymbol{\Sigma}}_{R}=\boldsymbol{D}^{1/2}_{n}\hat{\boldsymbol{R}}(\lambda_{n})\boldsymbol{D}^{1/2}_{n}$, where $\boldsymbol{D}_{n}={\rm diag}(\boldsymbol{\Sigma}_{n})$.
It is easy to see that a good choice of  the threshold $\lambda_{n}$  is $\lambda_{n}=C\sqrt{(\log p)/n}$ for some constant $C>0$. It is however difficult to choose $C$ because the choice depends on the unknown underlying distribution. Assuming the constant $C$ is chosen sufficiently large, it  can be shown that the resulting estimator $\hat{\boldsymbol{\Sigma}}_{R}$ attains the same minimax rate of convergence.  However, the estimator $\hat{\boldsymbol{\Sigma}}_{R}$ is less efficient than $\hat{\boldsymbol{\Sigma}}^{\star}(\delta)$ for support recovery. In fact, $\hat{\boldsymbol{\Sigma}}_{R}$  is unable to recover the support of $\boldsymbol{\Sigma}_{0}$ exactly for a class of non-Gaussian distributions of $\textbf{X}$.
Denote by $\mathcal{V}(\gamma,\delta, K_{1})$ the class of  distributions $F$ of $\textbf{X}$ satisfying the conditions of Theorem \ref{t2}. Then it can be shown that for any $\gamma>0$, $\delta\geq 2$ and some $K_{1}=K_{1}(\gamma)>0$,
 \beq
 \label{correlation.support}
\inf_{\lambda_{n}}\sup_{F\in \mathcal{V}(\gamma,\delta, K_{1})}\pr\Big{(}{\rm
supp}(\hat{\boldsymbol{\Sigma}}_{R})\neq{\rm
supp}(\boldsymbol{\Sigma}_{0})\Big{)}\rightarrow 1.
\eeq

The sample correlation coefficients $\hat r_{ij}$ are not homoscedastic,
although the range of variabilities is smaller in comparison to that of sample covariances. This is in fact the main reason for the negative result on support recovery given in Equation (\ref{correlation.support}). A natural approach to deal with the heteroscedasticity of the sample correlation coefficients is to first stabilize the variance by using Fisher's $z$-transformation, then threshold and finally obtain the estimator by inverse transform. Applying Fisher's $z$-transformation to each correlation coefficient yields
\[
\hat{Z}_{ij}={1\over 2}\ln {1+\hat{r}_{ij}\over 1-\hat{r}_{ij}}.
\]
When $\textbf{X}$ is multivariate normal, it is well-known that $\hat{Z}_{ij}$ is asymptotically normal with mean $(1/2)\ln((1+r_{ij})/(1-r_{ij}))$ and variance $1/(n-3)$. The behavior of $\hat{Z}_{ij}$ in the non-Gaussian case is more complicated. In general, the asymptotic variance of $\hat{Z}_{ij}$ depends on $\ep X^{2}_{i}X^{2}_{j}$ even when $r_{ij}=0$; see Hawkins (1989). Similar to the method of thresholding the sample correlation coefficients discussed earlier, universally thresholding  $(\hat{Z}_{ij})_{p\times p}$ is unable to recover the support of $\boldsymbol{\Sigma}_{0}$ exactly for a class of non-Gaussian distributions of $\textbf{X}$ satisfying the conditions in Theorem \ref{t2}.

In conclusion, the two natural approaches based on the sample correlation matrix discussed above are not as efficient as the entry-dependent thresholding method we proposed in Section \ref{method.sec}. For reasons of space, we omit the proofs of the results stated in this section. We shall explore these issues in detail elsewhere.

\section{Proofs}
\label{proof.sec}

We begin by collecting a few technical lemmas which are essential for
the proofs of the main results.
The first lemma is an exponential inequality on the partial sums of
independent random variables.

\begin{lemma}\label{ie1}
Let $\xi_{1},\cdots, \xi_{n}$ be independent random variables with mean
zero. Suppose that there exists some $t>0$ and $\bar{B}_{n}$ such
that $\sum_{k=1}^{n}\ep \xi^{2}_{k}e^{t|\xi_{k}|}\leq \bar{B}^{2}_{n}.$
Then for $0<x\leq \bar{B}_{n}$,
\begin{eqnarray}\label{eq1}
\pr\Big{(}\sum_{k=1}^{n}\xi_{k}\geq C_{t}\bar{B}_{n}x\Big{)}\leq
\exp(-x^{2}),
\end{eqnarray}
where $C_{t}=t+t^{-1}$.
\end{lemma}

\noindent{\bf Proof of Lemma \ref{ie1}.} By the inequality
$|e^{s}-1-s|\leq s^{2}e^{s\max(s,0)}$,  we have for any $t\geq 0$,
\begin{eqnarray*}
\pr\Big{(}\sum_{k=1}^{n}\xi_{k}\geq C_{K}\bar{B}_{n}x\Big{)}&\leq&
\exp(-tC_{\eta}\bar{B}_{n}x)\prod_{k=1}^{n}\ep\exp(t \xi_{k})\cr
&\leq& \exp(-tC_{\eta}\bar{B}_{n}x)\prod_{k=1}^{n}(1+t^{2}\ep
\xi^{2}_{k}e^{t|\xi_{k}|} )\cr &\leq&
\exp\Big{(}-tC_{\eta}\bar{B}_{n}x+\sum_{k=1}^{n}t^{2}\ep
\xi^{2}_{k}e^{t|\xi_{k}|}\Big{)}.
\end{eqnarray*}
Take $t=\eta(x/\bar{B}_{n})$. It follows that
\begin{eqnarray*}
\pr\Big{(}\sum_{k=1}^{n}\xi_{k}\geq C_{\eta}\bar{B}_{n}x\Big{)}\leq
\exp\Big{(}-\eta C_{\eta}x^{2}+\eta^{2}x^{2}\Big{)}=\exp(-x^{2}),
\end{eqnarray*}
which completes the proof.\qed\\

The second and third lemmas are on the asymptotic behaviors of the
largest entry of the sample covariance matrix and
$\hat{\theta}_{ij}$.   The proof of Lemma \ref{le2} is given in Cai and Liu (2010).

 \begin{lemma}\label{le2} (i). Under (C1), we have for any $\delta\geq
 2$, $\varepsilon>0$ and $M>0$,
 \begin{eqnarray}\label{lemma1}
 \pr\Big{(}\max_{ij}|\hat{\sigma}_{ij}-\sigma^{0}_{ij}|/\hat{\theta}^{1/2}_{ij}\geq
 \delta\sqrt{\log p/n}\Big{)}= O\Big{(}(\log
 p)^{-1/2}p^{-\delta+2}\Big{)},
 \end{eqnarray}
 \begin{eqnarray}\label{lemma1-1}
\pr\Big{(}\max_{ij}\{|\hat{\theta}_{ij}-\theta_{ij}|\}\geq
\varepsilon\sigma^{0}_{ii}\sigma^{0}_{jj}\Big{)}=O(p^{-M}),
 \end{eqnarray}
 and
 \begin{eqnarray}\label{lemma1-2}
\pr\Big{(}\max_{i}|\bar{X}^{i}|\geq C\sqrt{\log
p/n}\Big{)}=O(p^{-M})
\end{eqnarray}
 for some $C>0$.

(ii). Under (C2), (\ref{lemma1})-(\ref{lemma1-2}) still hold if we
replace $O\Big{(}(\log
 p)^{-1/2}p^{-\delta+2}\Big{)}$ and $O(p^{-M})$ with $O\Big{(}(\log
 p)^{-1/2}p^{-\delta+2}+n^{-\epsilon/8}\Big{)}$ and $O(n^{-\epsilon/8})$ respectively.
 \end{lemma}

\begin{lemma}\label{lesc} Let $\textbf{X}=(X_{1},\cdots,X_{p})$ be a mean zero
random vector. Suppose that $\Cov(\textbf{X})=I_{p\times p}$, (C3)
holds and $p\rightarrow\infty$. Then under (C1) or (C2), we have for
any $\delta>0$,
\begin{eqnarray*}
\pr\Big{(}\max_{1\leq i<j\leq
p}(n\theta_{ij})^{-1}\Big{|}\sum_{k=1}^{n}X_{ki}X_{kj}\Big{|}^{2}\geq
(4-\delta)\log p\Big{)}\rightarrow 1.
\end{eqnarray*}
\end{lemma}

\noindent{\bf Proof of Lemma \ref{lesc}.} We arrange the two dimensional
indices $\{(i,j):1\leq i<j\leq p\}$ in any ordering and set them as
$\{(i_{m},j_{m}): 1\leq m\leq p(p-1)/2=:L\}$. Let
\begin{eqnarray*}
Y_{km}=\theta^{-1/2}_{ij}X_{ki_{m}}X_{kj_{m}},\quad S_{m}=n^{-1/2}\sum_{k=1}^{n}Y_{km},\quad A_{m}=\{|S_{m}|\geq \sqrt{(4-\delta)\log p}\}, \quad
1\leq m\leq L.
\end{eqnarray*}
Define
$\bar{Y}_{km}=Y_{km}I\{|Y_{km}|\leq \delta_{n}\sqrt{n/(\log
p)^{3}}\}$ and $\hat{Y}_{km}=\bar{Y}_{km}-\ep \bar{Y}_{km}$,
where $\delta_{n}\rightarrow 0$ sufficiently slow. Then by (C1) or (C2) we have when $n$ is large,
\begin{eqnarray}\label{a37}
&&\pr\Big{(}\max_{1\leq i<j\leq p}(n\theta_{ij})^{-1}\Big{|}\sum_{k=1}^{n}X_{ki}X_{kj}\Big{|}^{2}\geq (4-\delta)\log p\Big{)}\cr &&\geq
\pr\Big{(}\max_{1\leq m\leq L}n^{-1}\Big{|}\sum_{k=1}^{n} \hat{Y}_{km}\Big{|}^{2}\geq (4-2\delta)\log p\Big{)}-O(p^{-M}+n^{-\epsilon/8})\cr &&\geq
\pr\Big{(}\max_{1\leq m\leq L}n^{-1}\Big{|}\sum_{k=1}^{n} \hat{Y}_{km}\Big{|}^{2}\geq 4\log p -\log\log p+x\Big{)}-O(p^{-M}+n^{-\epsilon/8}).
\end{eqnarray}
for any $M>0$ and $x<0$. Set $y_{n}=\sqrt{4\log p -\log\log p+x}$ and
\begin{eqnarray*}
\hat{A}_{m}=\Big{\{}n^{-1/2}\Big{|}\sum_{k=1}^{n}\hat{Y}_{km}\Big{|}\geq y_{n}\Big{\}}.
\end{eqnarray*}
 Then by Bonferroni's inequality, we have for
any fixed $l$,
\begin{eqnarray}\label{a36}
\pr\Big{(}\max_{1\leq m\leq L}n^{-1}\Big{|}\sum_{k=1}^{n} \hat{Y}_{km}\Big{|}^{2}\geq y^{2}_{n}\Big{)}\geq \sum_{d=1}^{2l}(-1)^{d-1}\sum_{1\leq
i_{1}<\cdots<i_{d}\leq L}\pr\Big{(}\bigcap_{j=1}^{d}\hat{A}_{i_{j}}\Big{)}.
\end{eqnarray}
Write
\begin{eqnarray*}
\hat{\textbf{Y}}_{k}=(\hat{Y}_{ki_{1}},\cdots,\hat{Y}_{ki_{d}}),\quad
1\leq k\leq n.
\end{eqnarray*}
 By Theorem 1 in Zaitsev (1987), we have
\begin{eqnarray}\label{a32}
&&\pr\Big{(}|\hat{\textbf{N}}|_{d,\infty}\geq
y_{n}-\delta^{1/2}_{n}(\log p)^{-1/2}\Big{)} +
c_{1}\exp(-c_{2}\delta^{-1/2}_{n}\log p)\cr
&&\geq\pr\Big{(}\Big{|}n^{-1/2}\sum_{k=1}^{n}\hat{\textbf{Y}}_{k}\Big{|}_{d,\infty}\geq
y_{n}\Big{)}\cr&&\geq \pr\Big{(}|\hat{\textbf{N}}|_{d,\infty}\geq
y_{n}+\delta^{1/2}_{n}(\log p)^{-1/2}\Big{)}-
c_{1}\exp(-c_{2}\delta^{-1/2}_{n}\log p),
\end{eqnarray}
where $c_{1}$ and $c_{2}$ are positive constant depending only on
$d$,  $|\cdot|_{d,\infty}$ means
$|\textbf{a}|_{d,\infty}=\min_{1\leq i\leq d}|a_{i}|$ for
$\textbf{a}=(a_{1},\cdots,a_{d})$, and $\hat{\textbf{N}}$ is a $d$
dimensional normal random vector with mean zero and covariance
matrix $\Cov(\hat{\textbf{Y}}_{k})$. Set
\begin{eqnarray*}
\hat{B}^{\pm}_{i_{1},\cdots,i_{d}}=\Big{\{}|\hat{\textbf{N}}|_{d,\infty}\geq
y_{n}\mp\delta^{1/2}_{n}(\log p)^{-1/2}\Big{\}}.
\end{eqnarray*}
 We can check that $\|\Cov(\hat{\textbf{N}_{k}})-I_{d\times
d}\|_{2}=O(1/(\log p)^{8})$. Let $\textbf{Z}$ be a standard
$d$-dimensional normal vector. Then we have
\begin{eqnarray}\label{a34}
\pr\Big{(}\hat{B}^{+}_{i_{1},\cdots,i_{d}}\Big{)}&\leq& \pr\Big{(}
|\textbf{Z}|_{d,\infty}\geq y_{n}-2\delta^{1/2}_{n}(\log
p)^{-1/2}\Big{)}\cr & &+\pr\Big{(}
\|\Cov(\hat{\textbf{N}_{k}})-I_{d\times d}\|_{2}|\textbf{Z}|_{2}\geq
\delta^{1/2}_{n}(\log p)^{-1/2}\Big{)}\cr
&=&(1+o(1))\Big{(}\frac{1}{\sqrt{2\pi}}p^{-2}\exp(-x/2)\Big{)}^{d}+O(\exp(-C(\log
p)^{2})).
\end{eqnarray}
Similarly we can get
\begin{eqnarray}\label{a35}
\pr\Big{(}\hat{B}^{-}_{i_{1},\cdots,i_{d}}\Big{)}\geq (1-o(1))\Big{(}\frac{1}{\sqrt{2\pi}}p^{-2}\exp(-x/2)\Big{)}^{d}-O(\exp(-C(\log p)^{2})).
\end{eqnarray}
Submitting (\ref{a32})-(\ref{a35}) into (\ref{a36}), we can get
\begin{eqnarray}\label{a38}
\liminf_{n\rightarrow\infty}\pr\Big{(}\max_{1\leq m\leq L}n^{-1}\Big{|}\sum_{k=1}^{n} \hat{Y}_{km}\Big{|}^{2}\geq y^{2}_{n}\Big{)}&\geq&
\sum_{d=1}^{2l}(-1)^{d-1}\Big{(}\frac{1}{\sqrt{8\pi}}\exp(-x/2)\Big{)}^{d}/d!\cr &\rightarrow
&1-\exp\Big{(}-\frac{1}{\sqrt{8\pi}}\exp(-x/2)\Big{)}
\end{eqnarray}
as $l\rightarrow\infty$. Letting $x\rightarrow-\infty$, we prove the
lemma by (\ref{a37}) and (\ref{a38}).\qed\\


\noindent{\bf Proof of Theorem \ref{bpsth4}.}   By (C1) or (C2),
 we have $\theta_{ij}\leq C_{K_{1}}\sigma^{0}_{ii}\sigma^{0}_{jj}$.  On the event
 $\{\max_{ij}|\hat{\sigma}_{ij}-\sigma^{0}_{ij}|\leq \lambda_{ij}\}\cap\{\hat{\theta}_{ij}\leq 2\theta_{ij}$ for all
 $i,j\}$, we have by the conditions (i)-(iii) on $s_{\lambda}(z)$
 that
\begin{eqnarray*}
&&\sum_{j=1}^{p}|s_{\lambda_{ij}}(\hat{\sigma}_{ij})-\sigma^{0}_{ij}|\cr
  &&= \sum_{j=1}^{p}|s_{\lambda_{ij}}(\hat{\sigma}_{ij})-\sigma^{0}_{ij}|I\{|\hat{\sigma}_{ij}|\geq
  \lambda_{ij}\}+\sum_{j=1}^{p}|\sigma^{0}_{ij}|I\{|\hat{\sigma}_{ij}|<
  \lambda_{ij}\}\cr
  &&\leq 2\sum_{j=1}^{p}\lambda_{ij}I\{|\sigma^{0}_{ij}|\geq
  \lambda_{ij}\}+\sum_{j=1}^{p}|s_{\lambda_{ij}}(\hat{\sigma}_{ij})-\sigma^{0}_{ij}|I\{|\hat{\sigma}_{ij}|\geq
  \lambda_{ij},|\sigma^{0}_{ij}|<
  \lambda_{ij}\}\cr
&&\quad+  \sum_{j=1}^{p}|\sigma^{0}_{ij}|I\{|\sigma^{0}_{ij}|<
  2\lambda_{ij}\}\cr
  &&\leq  2\sum_{j=1}^{p}\lambda^{1-q}_{ij}|\sigma^{0}_{ij}|^{q}+(1+c)\sum_{j=1}^{p}|\sigma^{0}_{ij}|I\{|\sigma^{0}_{ij}|<
  \lambda_{ij}\}+\sum_{j=1}^{p}|\sigma^{0}_{ij}|I\{|\sigma^{0}_{ij}|<
  2\lambda_{ij}\}\cr
  &&\leq
  C_{q,c}\sum_{j=1}^{p}\lambda^{1-q}_{ij}|\sigma^{0}_{ij}|^{q}\cr
  &&\leq C_{K_{1},\delta,c,q}s_{0}(p)\Big{(}\frac{\log
  p}{n}\Big{)}^{(1-q)/2}.
\end{eqnarray*}
The proof follows from Lemma \ref{le2} and the fact
$\|\boldsymbol{A}\|_{2}\leq \|\boldsymbol{A}\|_{L_{1}}$ for any symmetric matrix.\qed
\\

\noindent{\bf Proof of Theorems 2 and \ref{th4}.}  Theorem 2 follows
from Lemma \ref{le2} immediately. We now prove Theorem \ref{th4}. For each $1\leq i\leq p$, let
$A_{1}$ be the largest subset of $\{1,\cdots,p\}$ such that $X_{i}$
is uncorrelated with $\{X_{k},k\in A_{1}\}$. Let
$i_{1}=$argmin$\{|j-i|:j\in A_{1}\}$.
 Then we have $|i_{1}-i|\leq s$. Also, Card$(A_{1})\geq p-s$.
Similarly, let $A_{l}$ be the largest subset of $A_{l-1}$ such that
$X_{i_{l-1}}$ is uncorrelated with $\{X_{k},k\in A_{l}\}$ and
$i_{l}=$argmin$\{|j-i_{l-1}|: j\in A_{l}\}$.  We can see that
$|i_{l}-i|\leq ls$ and Card$(A_{l})\geq$Card$(A_{l-1})-s\geq p-sl$.
Take $l=[p^{\tau_{2}}]$ with
$\tau^{2}/4<\tau_{2}<\min(\tau^{2}/3,\tau_{1})$.
 Then
$X_{i_{0}},\ldots,X_{i_{l}}$ are pairwise uncorrelated random
variables, where we set $i_{0}=i$. Clearly $i_{1},\cdots,i_{l}\in
B_{i}=\{j: \sigma^{0}_{ij}=0; j\neq i\}$.
 Without loss of
generality, we assume that $X_{1},\cdots, X_{l}$ are pairwise
uncorrelated. Note that $|s_{\lambda}(z)|\geq |z|-\lambda$. It
suffices to show that for some $\varepsilon_{0}>0$,
\begin{eqnarray}\label{a5}
\pr\Big{(}\max_{1\leq i<j\leq
l}\{\lambda_{nij}^{-1}|\hat{\sigma}_{ij}|\}> 1+\varepsilon_{0}
\Big{)}\rightarrow 1.
\end{eqnarray}
 Clearly, we can assume $\ep \textbf{X}=0$ and
$\Var(X_{i})=1$ for $1\leq i\leq l$. By Lemma \ref{le2} and
(\ref{c2}), we have $\min_{ij}\lambda_{nij}>0$ with probability
tending to one.
 By Lemma \ref{le2} it suffices to show that
 for any $0<\tau<2$,
 \begin{eqnarray}\label{a6}
A_{n}:=\pr\Big{(}\max_{1\leq i<j\leq
l}\Big{\{}(n\theta_{ij})^{-1/2}\Big{|}\sum_{k=1}^{n}X_{ki}X_{kj}\Big{|}\Big{\}}\geq
\tau\sqrt{\log p}\Big{)}\rightarrow 1.
 \end{eqnarray}
Since $\tau^{2}\log p\leq (4-\delta)\log l $ for
$0<\delta<4-\tau^{2}/\tau_{2}$ and large $n$, (\ref{a6}) follows
from Lemma 3.
 \qed\\

Lemmas \ref{le4} and \ref{le5} below, proved in Cai and Liu (2010), are needed to
prove Theorems \ref{th5} and \ref{th6}.
\begin{lemma}
\label{le4}
Suppose that $\textbf{X}\sim N(\boldsymbol{\mu},\boldsymbol{\Sigma}_{0})$ with $\boldsymbol{\Sigma}_{0}\in\bar{\mathcal{U}}_{0}$. Let $s_{0}(p)=O((\log p)^{\gamma})$ for some
$\gamma<1$ and $n^{\xi}\leq p\leq \exp(o(n^{1/3}))$ for some
$\xi>0$. Let $\delta>\sqrt{2}$. Then there are at most $O(s_{0}(p))$
nonzero elements in each row of $\hat{\boldsymbol{\Sigma}}^{\star}(\delta)$.
Furthermore,
\begin{eqnarray}
\label{a40}
\inf_{\boldsymbol{\Sigma}_{0}\in\bar{\mathcal{U}}_{0}}\pr\Big{(}\|\hat{\boldsymbol{\Sigma}}^{\star}(\delta)-\boldsymbol{\Sigma}_{0}\|_{2}\leq
C_{\gamma,\delta,M}\max_{i}\sigma^{0}_{ii}s_{0}(p)\Big{(}\frac{\log
p}{n}\Big{)}^{1/2}\Big{)}\geq 1-O(p^{-M})
\end{eqnarray}
for any $M>0$, where $C_{\gamma,\delta,M}$ is a constant depending
only on $\gamma,\delta,M$, and
\begin{eqnarray}\label{a41}
\sup_{\boldsymbol{\Sigma}_{0}\in\mathcal{U}_{0}}\ep\|\hat{\boldsymbol{\Sigma}}^{\star}(\delta)-\boldsymbol{\Sigma}_{0}\|^{2}_{2}\leq
Cs^{2}_{0}(p)\frac{\log p}{n}
\end{eqnarray}
for some constant $C>0$.
 \end{lemma}

\begin{lemma}
\label{le5}
Let  $\lambda_{ij}=\tau\sqrt{\frac{\hat{\theta}_{ij}\log
 p}{n}}$ with $0<\tau<\sqrt{2}$.  Under the conditions of Lemma
4,
\begin{eqnarray}\label{a19}
\pr\Big{(}\min_{i}\sum_{j\in B_{i}}I\{|\hat{\sigma}_{ij}|\geq
\lambda_{nij}(\tau)\}\geq p^{2\epsilon_{0}}\Big{)}\rightarrow 1
\end{eqnarray}
with any $\epsilon_{0}<(1-\tau^{2}/2)/2$, where $B_{i}=\{j:
\sigma^{0}_{ij}=0; j\neq i\}$. Hence for some constant $C>0$,
\begin{eqnarray*}
\inf_{\boldsymbol{\Sigma}_{0}\in
\bar{\mathcal{U}}_{0}}\pr\Big{(}\|\hat{\boldsymbol{\Sigma}}^{\star}(\tau)-\boldsymbol{\Sigma}_{0}\|_{2}\geq
C\min_{i}\sigma^{0}_{ii}p^{\epsilon_{0}/2}s_{0}(p)\Big{(}\frac{\log
p}{n}\Big{)}^{1/2}\Big{)}\rightarrow 1.
\end{eqnarray*}
\end{lemma}

\noindent{\bf Proof of Theorem \ref{th5}.} To simplify the notation, we shall write $s_0$ for $s_{0}(p)$. We construct a matrix
$\boldsymbol{\Sigma}_{0}\in\mathcal{U}^{\star}_{q}$.  Let
$s_{1}=[(s_{0}-1)^{1-q}(\log p/n)^{-q/2}]+1$ and
$(X_{1},\cdots,X_{s_{1}})$, $X_{s_{1}+1},\cdots,X_{p}$ be
independent. Let $\sigma^{0}_{ii}=s_{0}$ for all $i>s_{1}$,
$\sigma^{0}_{ii}=1$ for $1\leq i\leq s_{1}$ and
$\sigma^{0}_{ij}=4^{-1}s_{0}\sqrt{\log p/n}$ for $1\leq i\neq j\leq
s_{1}$. Note that $\sigma^{0}_{ij}=0$ for $i\neq j>s_{1}$. Since
$s_{0}<4\sqrt{n/\log p}$, $\boldsymbol{\Sigma}_{0}$ is a positive definite
covariance matrix belonging to $\mathcal{U}^{\star}_{q}$. Set
$\boldsymbol{M}_{n}=(\sigma^{0}_{ij})_{1\leq i,j\leq s_{1}}$. We first suppose
that $\lambda_{n}\leq 3^{-1}\sigma^{0}_{pp}\sqrt{2\log p/n}$.
Lemma \ref{le5} yields
\begin{eqnarray*}
\pr\Big{(}\sum_{j=s_{1}+1}^{p}I\{|\hat{\sigma}_{pj}|\geq
\frac{\sqrt{2}}{2}\sigma^{0}_{pp}\sqrt{\frac{\log p}{n}}\}\geq
p^{2\epsilon_{0}}\Big{)}\rightarrow 1,
\end{eqnarray*}
with any $\epsilon_{0}<3/8$. Take $\epsilon_{0}=7/20$ and note that
$p^{1/4}\geq s_{0}$, $p^{1/10}\geq n^{q/2}$. By the inequality
$|s_{\lambda}(z)|\geq z-\lambda$,
\begin{eqnarray}\label{a28}
\inf_{\lambda_{n}\leq3^{-1}\sigma^{0}_{pp}\sqrt{2\log p/n}
}\sup_{\mathcal{U}^{\star}_{q}}\pr\Big{(}\|\hat{\boldsymbol{\Sigma}}_{g}-\boldsymbol{\Sigma}_{0}\|_{2}>
\frac{\sqrt{2}}{6}s^{2}_{0}(p)\Big{(}\frac{\log
p}{n}\Big{)}^{(1-q)/2}\Big{)}\rightarrow 1.
\end{eqnarray}
We next consider the case
$\lambda_{n}>3^{-1}\sigma^{0}_{pp}\sqrt{2\log p/n}$. We have
\begin{eqnarray*}
\|\hat{\boldsymbol{\Sigma}}_{g}-\boldsymbol{\Sigma}_{0}\|_{2}\geq \|\hat{\boldsymbol{M}}_{n}-\boldsymbol{M}_{n}\|_{2},
\end{eqnarray*}
where $\hat{\boldsymbol{M}}_{n}=(\hat{\sigma}^{g}_{ij})_{1\leq i,j\leq s_{1}}$.
As in Lemma 2, we can get for any $\gamma>0$
\begin{eqnarray*}
\pr\Big{(}\max_{1\leq i,j\leq
s_{1}}|\hat{\sigma}_{ij}-\sigma^{0}_{ij}|\geq \sqrt{2\gamma\log
p/n}\Big{)}\leq Cs^{2}_{1}(\log p)^{-1/2}p^{-\gamma}.
\end{eqnarray*}
Taking $\gamma=1$,  we have with probability tending to one,
$\max_{1<i<j\leq s_{1}}|\hat{\sigma}_{ij}|\leq
(4^{-1}s_{0}+\sqrt{2})\sqrt{\log p/n}$, which implies that
 $\hat{\sigma}^{g}_{ij}=0$ for $1\leq i\neq j\leq s_{1}$. Thus, with probability tending to one,
 \begin{eqnarray*}
\|\hat{\boldsymbol{M}}_{n}-\boldsymbol{M}_{n}\|_{2}\geq
(4^{-1}-\sqrt{2}s_{0}^{-1})s_{1}s_{0}\sqrt{\frac{\log p}{n}}\geq
\frac{3}{64}s^{2-q}_{0}\Big{(}\frac{\log p}{n}\Big{)}^{(1-q)/2}.
 \end{eqnarray*}
 This and (\ref{a28}) together imply (\ref{ap1}). \qed
\\

\noindent{\bf Proof of Theorem \ref{th6} and Proposition
\ref{prop2}.} For brevity, we only consider the case $H=1$. The
proof for general $H$ is similar.
 We first show that for any $\varepsilon>0$,
\begin{eqnarray}\label{a24}
\pr\Big{(}\hat{\delta}\geq \sqrt{2}-\varepsilon\Big{)}\rightarrow 1 .
\end{eqnarray}
 Since
the random split is independent with the sample
$\{\textbf{X}_{1},\cdots,\textbf{X}_{n}\}$, we can assume that the
two samples are $\{\textbf{X}_{1},\cdots,\textbf{X}_{n_{1}}\}$ and
$\{\textbf{X}_{n_{1}+1},\cdots,\textbf{X}_{n}\}$. Let
$\hat{\boldsymbol{\Sigma}}_{2}$ be the sample covariance matrix from
$\{\textbf{X}_{n_{1}+1},\cdots,\textbf{X}_{n}\}$ and
$\hat{\boldsymbol{\Sigma}}^{\star}_{1}(\delta)$ be defined as in (\ref{adap})
from $\{\textbf{X}_{1},\cdots,\textbf{X}_{n_{1}}\}$. Define
\begin{eqnarray*}
\hat{\delta}_{o}=\hat{j}_{o}/N,\quad\mbox{where\quad}\hat{j}_{o}=\argmin_{0\leq
j\leq 4N}\|\hat{\boldsymbol{\Sigma}}^{\star}_{1}(j/N)-\boldsymbol{\Sigma}_{0}\|^{2}_{F}.
\end{eqnarray*}
Set
$a_{n}=p^{-1}\|\hat{\boldsymbol{\Sigma}}^{\star}_{1}(\hat{\delta})-\boldsymbol{\Sigma}_{0}\|^{2}_{F}$
and
$r_{n}=p^{-1}\|\hat{\boldsymbol{\Sigma}}^{\star}_{1}(\hat{\delta}_{o})-\boldsymbol{\Sigma}_{0}\|^{2}_{F}$.
By the proof of Theorem 1, we have
$\pr\Big{(}\|\hat{\boldsymbol{\Sigma}}^{\star}_{1}(2)-\boldsymbol{\Sigma}_{0}\|_{L_{1}}\leq
C_{1}s_{0}(p)(\log p/n)^{1/2}\Big{)}\rightarrow 1$ for some
$C_{1}>0$. Using the inequality $p^{-1}\|\boldsymbol{A}\|^{2}_{F}\leq
|\boldsymbol{A}|_{\infty}\|\boldsymbol{A}\|_{L_{1}}$ for any $p\times p$ symmetric matrix $\boldsymbol{A}$ and
the definition of $\hat{\delta}_{o}$, we have
$\pr\Big{(}r_{n}\leq C_{2}s_{0}(p)\log p/n\Big{)}\rightarrow 1$ for
some $C_{2}>0$. Note that
\begin{eqnarray*}
\ep|(\boldsymbol{V},\hat{\boldsymbol{\Sigma}}_{2}-\boldsymbol{\Sigma}_{0})|^{2}\leq Cn^{-1}
\end{eqnarray*}
for any $p\times 1$ vector $\boldsymbol{V}$ with $\|\boldsymbol{V}\|_{F}=1$. By the proof of
Theorem 3 in Bickel and Levina (2008) and the assumption that $N$
is fixed, we can see that,
\begin{eqnarray}\label{a23}
a_{n}\leq
O_{\pr}\Big{(}\frac{1}{n^{1/2}}\Big{)}a^{1/2}_{n}+O_{\pr}\Big{(}\frac{1}{n^{1/2}}\Big{)}r^{1/2}_{n}+r_{n}.
\end{eqnarray}
Hence for some $C_{3}>0$,
\begin{eqnarray}\label{a22}
\pr\Big{(}a_{n}\leq C_{3}s_{0}(p)\log p/n\Big{)}\rightarrow 1.
\end{eqnarray}
 Note that by applying Lemma 5 to the samples $\{\textbf{X}_{1},\cdots,\textbf{X}_{n_{1}}\}$,
\begin{eqnarray*}
\pr\Big{(}a_{n}\leq C_{3}s_{0}(p)\log p/n,\hat{\delta}<
\sqrt{2}-\varepsilon\Big{)}=o(1).
\end{eqnarray*}
 This together with (\ref{a22}) shows that
\begin{eqnarray*}
\pr\Big{(}\hat{\delta}< \sqrt{2}-\varepsilon\Big{)}\leq
\pr\Big{(}\hat{\delta}< \sqrt{2}-\varepsilon,a_{n}\leq
C_{3}s_{0}(p)\log p/n\Big{)}+o(1)=o(1),
\end{eqnarray*}
and hence (\ref{a24}) holds. Since $N$ is fixed, we have
$|\hat{\sigma}-\sqrt{2}|\geq \varepsilon_{0}$ for some fixed
$\varepsilon_{0}>0$ which depends on $N$. This together with
(\ref{a24}) implies
\begin{eqnarray}\label{a43}
\pr\Big{(}\hat{\delta}\geq \sqrt{2}+\epsilon\Big{)}\rightarrow 1
\end{eqnarray}
for some $\epsilon>0$. By Lemma 4, we see that with probability
tending to one, for each $i$, there are at most $O(s_{0}(p))$
nonzero numbers of $\{|s_{\lambda_{ij}}(\hat{\sigma}_{ij})|; j\in
B_{i}\}$ and by Lemma 2, they are of order
$O(\max_{i}\sigma^{0}_{ii}\sqrt{\log p/n})$.  Let $\Psi_{i}=\{j:
\sigma^{0}_{ij}\neq 0\}$ and $\hat{\Psi}_{i}=\{j:
\hat{\sigma}^{\star}_{ij}\neq 0\}$. Then by the conditions on
$s_{\lambda}(z)$, we have
\begin{eqnarray}\label{a14}
 \|\hat{\boldsymbol{\Sigma}}^{\star}(\hat{\delta})-\boldsymbol{\Sigma}_{0}\|_{L_{1}}
\leq\max_{i}\sum_{j\in \Psi_{i}\cup
\hat{\Psi}_{i}}|s_{\lambda_{ij}}(\hat{\sigma}_{ij})-\sigma^{0}_{ij}|
\leq C\max_{i}\sigma^{0}_{ii}s_{0}(p)\Big{(}\frac{\log
p}{n}\Big{)}^{1/2}
\end{eqnarray}
with probability tending to one. The proof of Theorem \ref{th6} is
completed. Finally, Proposition \ref{prop2} is proved by
(\ref{a43}), Lemmas 2 and 4. \qed

\section{Supplemental materials}

\begin{description}

\item[Additional proofs:] {A supplement to the main paper contains additional technical arguments including the proofs of Lemmas 2, 4 and 5. (pdf file)}
\end{description}

\section*{Acknowledgment}

We  thank the Associate Editor and two referees for their helpful comments which have led to an improvement of the presentation of the paper.

\end{document}